\documentclass[preprint2]{aastex}

\shorttitle{Formation of sunspots and active regions, and origin of their asymmetries}
\shortauthors{Chen et al.}

\usepackage{natbib}
\usepackage{amsmath}
\usepackage{graphicx,subfigure}
\usepackage[percent]{overpic}
\usepackage{color}

\newcommand{\sectref}[1]{Section\,\ref{#1}}

\newcommand{\figref}[1]{Figure\,\ref{#1}}

\newcommand{\tabref}[1]{Table\,\ref{#1}}
\newcommand{\equref}[1]{Equation\,(\ref{#1})}

\begin{document}


\title{ Emergence of magnetic flux generated in a solar convective dynamo. \\
I: Formation of Sunspots and Active regions, and Origin of Their Asymmetries}


\author{Feng Chen\altaffilmark{1}, Matthias Rempel, and Yuhong Fan}
\affil{High Altitude Observatory, NCAR, P.O. Box 3000, Boulder, CO, 80307, USA}
\altaffiltext{1}{chenfeng@ucar.edu}


\begin{abstract}
We present a realistic numerical model of sunspot and active region formation based on the emergence of flux bundles generated in a solar convective dynamo. To this end we use the magnetic and velocity fields in a horizontal layer near the top boundary of the solar convective dynamo simulation to drive realistic radiative-magnetohydrodynamic simulations of the upper most layers of the convection zone. The main results are: (1) The emerging flux bundles rise with the mean speed of convective upflows, and fragment into small-scale magnetic elements that further rise to the photosphere, where bipolar sunspot pairs are formed through the coalescence of the small-scale magnetic elements. (2) Filamentary penumbral structures form when the sunspot is still growing through ongoing flux emergence. In contrast to the classical Evershed effect, the inflow seems to prevail over the outflow in a large part of the penumbra. (3) A well formed sunspot is a mostly monolithic magnetic structures that is anchored in a persistent deep-seated downdraft lane. The flow field outside the spot shows a giant vortex ring that comprises of an inflow below 15\,Mm depth and an outflow above 15\,Mm depth. (4) The sunspots successfully reproduce the fundamental properties of the observed solar active regions, including the more coherent leading spots with a stronger field strength, and the correct tilts of bipolar sunspot pairs. These asymmetries can be linked to the intrinsic asymmetries in the magnetic and flow fields adapted from the convective dynamo simulation.
\end{abstract}


\section{Introduction} \label{sec:intro}

Sunspots are observed as dark spots in the photosphere of the Sun with magnetic field of a few kilo Gauss that can significantly suppress convection. The umbra of a sunspot is the dark core area where the magnetic field is strongest and mostly vertical. The magnetic field lines become strongly inclined at the outer rim of the umbra and give rise to the numerous radial filamentary structures known as the penumbra \citep[see reviews by][]{Solanki:2003,Schlichenmaier:2009,Borrero+Ichimoto:2011}. Bipolar sunspot pairs, which are a very common category, appear as two magnetic flux concentrations of opposite polarities. The spot leading in the direction of the solar rotation (i.e., the {\it leading} or {\it proceeding} spot) is closer to the equator of the Sun than the {\it following} (or {\it trailing}) spot, and the tilt angles of the axes connecting the sunspot pairs are found to depend on the latitude of the sunspots. This is the well-known Joy's Law discovered by \citet{Hale+al:1919,Hale+Nicholson:1925} and confirmed by modern observations \citep[e.g.,][]{Wang+Sheeley:1989,Howard:1991,Stenflo+Kosovichev:2012}. The leading spots are also found to be more coherent than the following spots, i.e., the flux of the leading polarity is mostly concentrated in large and coherent magnetic structures, whereas the flux of the following polarity tends to be distributed in more fragmented structures \citep{Bray+Loughhead:1979}.

The magnetic field of the Sun is generated by the dynamo in the convection zone, albeit the detailed mechanism of the dynamo is still unclear and under debate \citep[see reviews by][]{Charbonneau:2005,Charbonneau:2010,Charbonneau:2014,Brun+al:2015}. Magnetic flux present in the convection zone is suggested to become unstable, buoyantly rise to the surface of the Sun, and finally gives rise to a pair of sunspots \citep{Parker:1955}. This scenario is strongly supported by the observational fact that the formation of the sunspots and solar active region is closely related to magnetic flux emergence \citep[e.g.,][]{Zwann:1985}. While direct observations are usually limited to the photosphere and layers above, helioseismology can shed some light on the flow pattern around and beneath magnetic flux concentrations \citep[e.g.,][and references therein]{Zhao+Kosovichev:2001,Gizon+Birch:2005}. This technique was also employed  to detect emerging magnetic structures in the convection zone \citep[e.g.,][]{IIonidis+al:2011,Birch+al:2013,Schunker+al:2016}. However, it gives very limited information on the magnetic field. The study of magnetic flux emergence, as well as the generation and intensification of magnetic flux, has to rely mostly on theoretical calculations.

Earlier works studied the rise of a toroidal thin flux tube from the base of the convection zone \citep{DSilva+Choudhuri:1993, Fan+al:1993, Caligari+al:1995} and found that the Coriolis force can produce tilt angles well consistent with observations. \citet{Fan+al:1993} and \citet{Caligari+al:1995} also showed that the flux tube becomes asymmetric between its leading (toward the direction of rotation) and following sides. \citet{Fan:2008} extended the one dimensional thin flux tube models (e.g., the aforementioned ones) into full three-dimensional models and considered the evolution of flux tubes in a marginally convectively-stable shell representing the solar convection zone. The author demonstrated that the leading side of the flux tube has stronger magnetic field than the following side due to asymmetric stretching of Coriolis force, and that the twist of the flux tube plays a crucial role in determining the tilt of the flux tube. The properties of the emerging flux tube considered in these models could explain those observed in active regions in the photosphere, if these properties can be retained when the flux tube reaches the photosphere. One important ingredient that is still missing in those models is the convective motion.  \citet{Jouve+Brun:2009} studied the rise of flux tubes through a spherical shell of turbulent convection. They found that in the case of a weak field the shape of the rising flux tube is dominated by convective up and down flows, leading to a flux tube that emerges in an $\Omega$ shape. \citet{Weber+al:2011} presented a one-dimensional flux tube model, in which the flow field is adapted from a global simulation of a rotating, turbulent convective envelope \citep{Miesch+al:2008}. They also found that an $\Omega$ shape structure forms when the magnetic field strength is relatively weak and the impact of the convective flow becomes significant.

3D magnetohydrodynamics simulations of the convective dynamo are a more self-consistent approach to study the generation and evolution of magnetic field. While recent simulations are able to produce a cyclic magnetic field in a rotating convective envelope \citep{Ghizaru+al:2010,Racine+al:2011,Nelson+al:2011, Nelson+al:2013, Nelson+al:2014, Fan+Fang:2014, Augustson+al:2015, Kaepylae+al:2012, Hotta+al:2016}, the detailed properties and parameters of the Sun's magnetic cycle are not fully reproduced yet. \citet{Nelson+al:2013} and \citet{Fan+Fang:2014} found that (toroidal) magnetic flux bundles\footnote{In this paper, "flux tube" refers to a magnetic flux system well isolated from surrounding magnetic structure and usually defined analytically. In contrast, "flux bundle" refers to distinctive flux system that might still have connections with other magnetic structures.} are formed in the turbulent convection zone and emerge due to buoyancy and advection by convective upflows. These models do not require a strong toriodal magnetic field at the base of the convection zone, and shed new light on the origin of magnetic structures that eventually give rise to solar active region. In particular, \citet{Fan+Fang:2014} found that the emerging flux bundle moves prograde in a giant convective cell, and hence is pushed closer to the downdraft lane at the leading side of the convection cell. Such an asymmetric pattern in the flow field leads to the stronger magnetic field at the leading side of the emerging flux bundle. Consequently, the solar active regions formed in the photosphere may reflect the same asymmetry.

However, it is not straight forward to project the properties of the sunspots at the surface from that of the emerging flux bundles, because the global dynamo simulations typically apply the anelastic approximation, which is not valid in the upper most convection zone. Moreover, most global dynamo simulations cover a density contrast of at most 100, i.e., a density contrast of about 10$^4$ is still missing above the top boundary. The emergence of magnetic flux through the upper most layer of the convection zone and further into the solar atmosphere needs to be investigated by fully compressible MHD simulations \citep[e.g.,][]{Fan:2001,Manchester+al:2004,Archontis+al:2004,Hood+al:2009}. Please refer to \citet{Fisher+al:2000}, \citet{Fan:2004,Fan:2009} and \citet{Cheung+Isobe:2014} for more comprehensive reviews on previous works. 

An other line of research has focused on modeling the interaction between convective flows, radiation effects and magnetic field in the photosphere by either using semi-empirical approximations for radiative source terms \citep[e.g.,][]{Abbett:2007,Fang+al:2010,Fang+al:2012} or by solving the radiative transfer equation \citep[e.g.,][]{Cheung+al:2007,Cheung+al:2008,Martinez+al:2008,Tortosa+Moreno:2009}.  While previous works mostly focused on the formation of the magnetic structures that are more similar to pores, \citet{Cheung+al:2010} simulated the emergence of a coherent torus shaped flux tube containing about $7\times10^{21}$\,Mx flux. They found that the initially coherent flux tube does not emerge as such into the photosphere, but fragments into small magnetic elements due to near surface convection. The small magnetic elements emerge into photosphere and coalesce to give rise to large sunspots. \citet{Rempel+Cheung:2014} extended this model by including a retrograde flow in the flux tube, which is motivated by the result from previous thin flux tube models, and found that the two sunspots formed in the photosphere are clearly asymmetric in morphology, as well as temporal evolution.

The work presented in this paper is a further extension of the works by \citet{Cheung+al:2010} and \citet{Rempel+Cheung:2014}. Instead of using an idealized semi-torus shaped flux tube that is advected by a uniform and artificial upflow through the bottom boundary, we employ the magnetic and flow fields extracted from a spherical surface near the top boundary of the global convective dynamo simulation described in \citet{Fan+Fang:2014}. Our aim is to study the evolution of the emerging flux bundles generated in this convective dynamo simulation through the upper most layer of the convection zone and the formation of sunspots and active regions in the photosphere. In this particular setup,  the magnetic and flow fields are dynamically consistent. This allows the simulations to produce not only more realistic sunspots and active regions in the photosphere, but also a more realistic flow field associated with the emerging magnetic structures. Moreover, because the emerging flux bundles already possess asymmetries similar to those observed in solar active regions at the surface, this model offers a unique opportunity to study if these properties can be preserved when the flux bundles emerge to the surface, and make direct comparisons with sunspot asymmetries in observations.

The rest of the paper is structured as follows. \sectref{sec:setup} describes the properties of the emerging flux bundles found in \citet{Fan+Fang:2014} and the detailed setup of the flux emergence simulation. \sectref{sec:res} presents the evolution of the magnetic field in the near-surface layer of the convection zone and the formation of sunspots and active regions at the photosphere. \sectref{sec:asym} presents the asymmetries in the sunspots and their origin. The implications of the results are discussed in \sectref{sec:sum}.

\section{Model setup} \label{sec:setup}
\subsection{Simulation of the Solar Convective Dynamo}\label{sec:setup_ff}

\begin{figure*}
\centering
\includegraphics[width=0.32\textwidth]{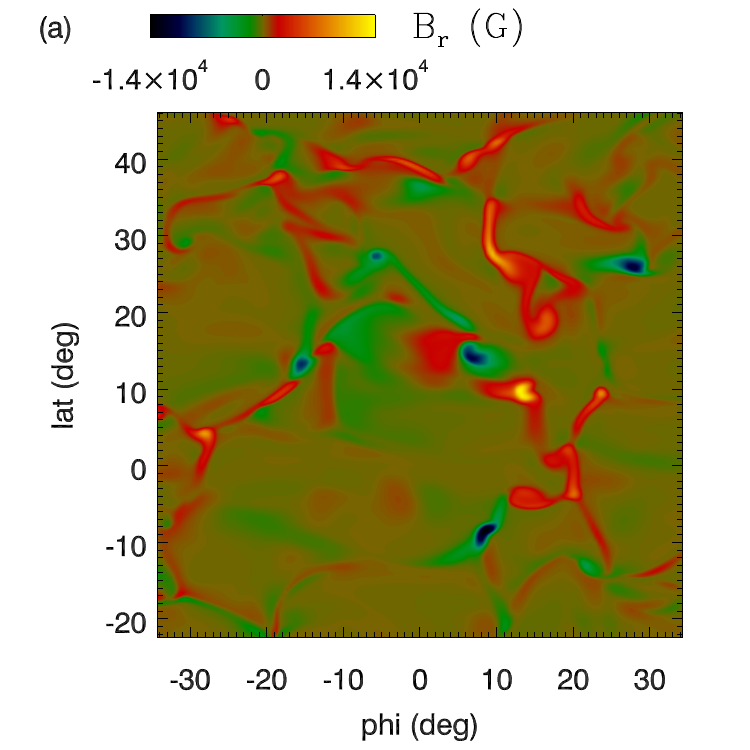}
\includegraphics[width=0.32\textwidth]{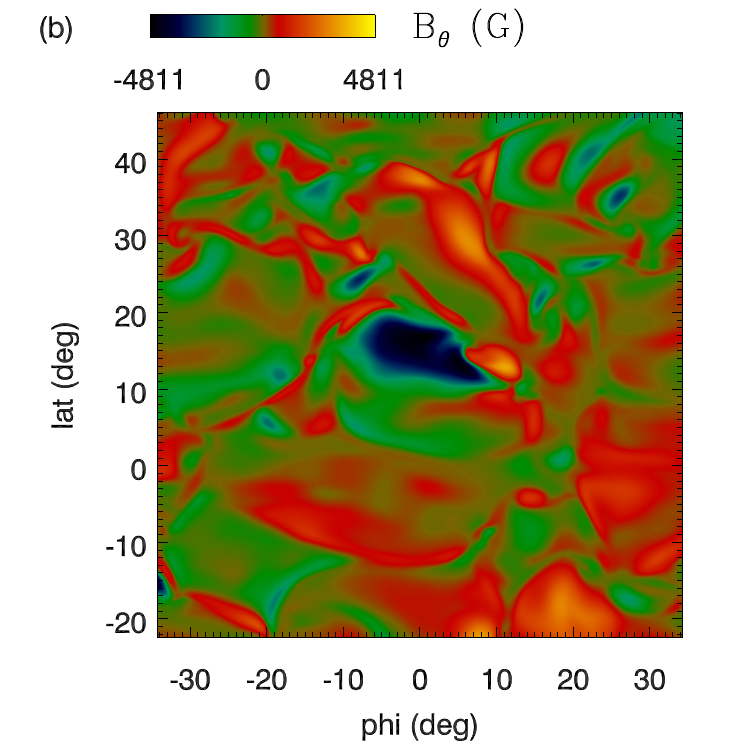}
\includegraphics[width=0.32\textwidth]{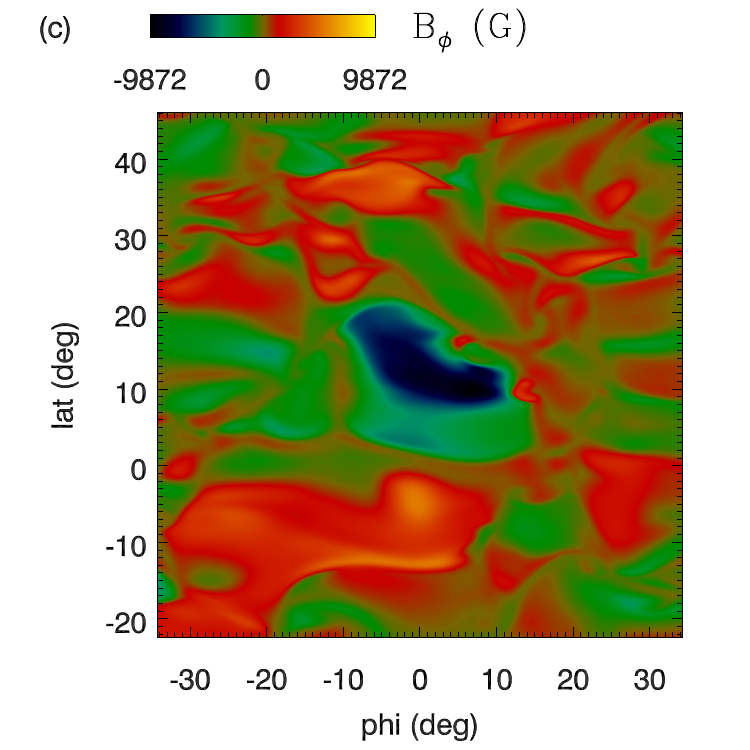} \\
\includegraphics[width=0.32\textwidth]{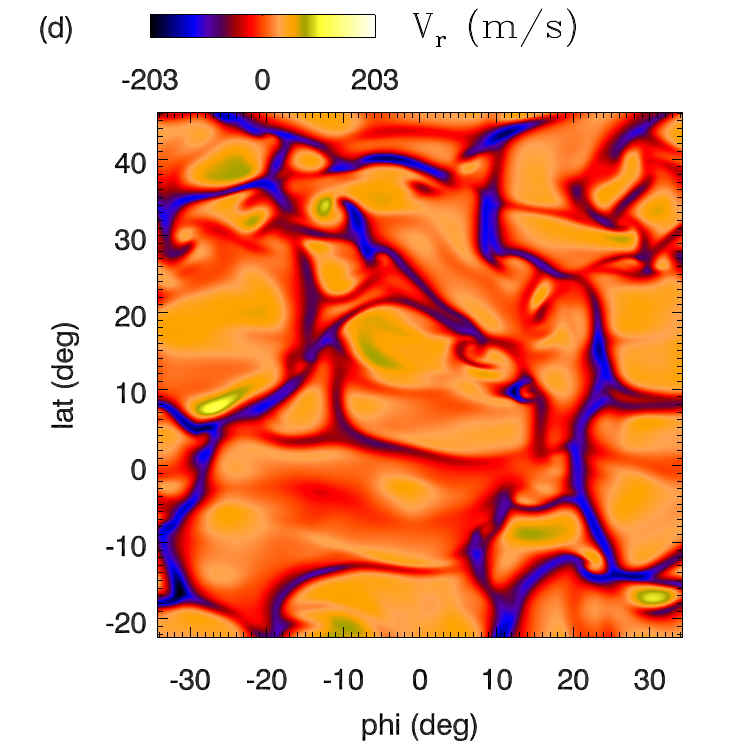}
\includegraphics[width=0.32\textwidth]{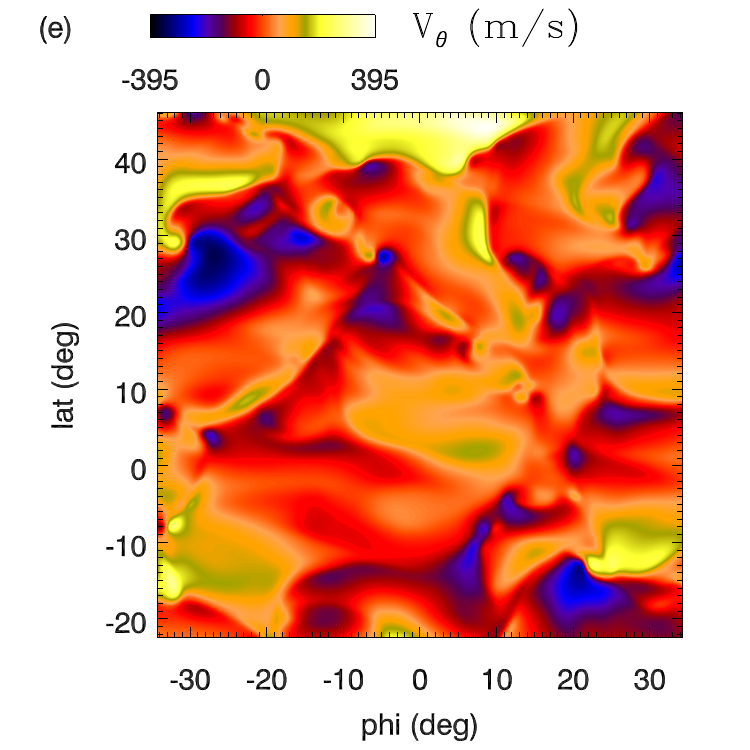}
\includegraphics[width=0.32\textwidth]{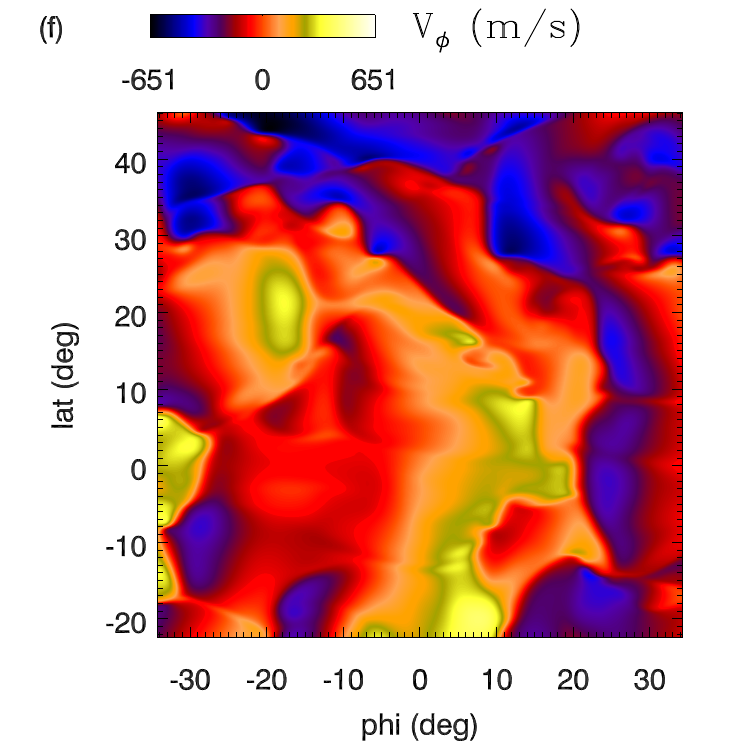}
\caption{Example of extracted slices of ${\bf B}$ and ${\bf v}$ field
at 30 Mm depth, corresponding to the time instance shown in Figure 10
of \citet{Fan+Fang:2014}. An animation is available in the online version of the paper.}
\label{fig:bvslices}
\end{figure*}

As described in \citet{Fan+Fang:2014}, the convective dynamo is driven by the solar radiative diffusive heat flux in a spherical shell rotating with the solar rotation rate. It produces a large-scale mean field that exhibits an irregular cyclic behavior with polarity reversals, and self-consistently maintains a solar-like differential
rotation. The convective dynamo also shows the emergence of strong super-equipartition flux bundles near the top of the simulation domain (located at 20 Mm below the photosphere). The flux bundles exhibit tilt angles that have a systematic mean consistent with the mean tilt of solar active regions \citep[see Figure 13 in][]{Fan+Fang:2014}.

\begin{figure*}
\includegraphics[width=0.96\textwidth]{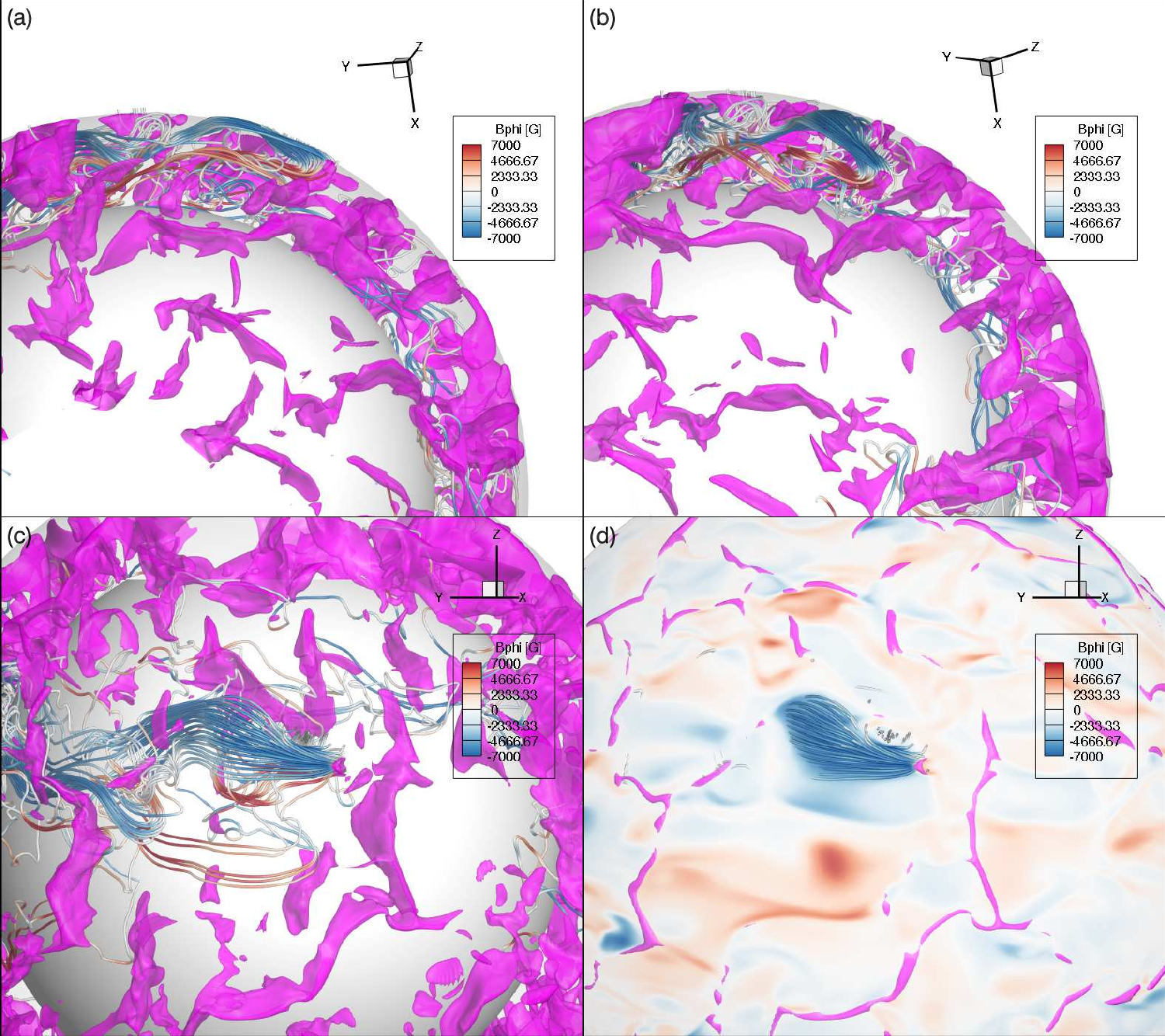}
\caption{(a)(b)(c) show selected field lines of the emerging flux bundle colored by $B_{\phi}$, and (pink) isosurfaces of strong downflows with $v_r$ reaching -110 m\,s$^{-1}$, viewed from 3 different perspectives. (d) shows the same perspective as (c) but with the spherical surface of $B_{\phi}$ at the depth of the extracted slices in \figref{fig:bvslices} included.\label{fig:emg3d}} 
\end{figure*}

We extract horizontal slices of ${\bf B}$ and ${\bf v}$ field at the depth of $30$ Mm below the photosphere, i.e., 10\,Mm below the top boundary of the simulation domain. The region spans about $69$ degree in both latitude and azimuth, and is centered on an emerging flux region that is shown in Figure 10 of \citet{Fan+Fang:2014} (marked by an arrow) at $12$ deg latitude. This region actually covers parts of both the northern and southern hemisphere, with about 46 deg in the former and 23 deg in the latter.

Moreover, the selected region moves with a local rotation rate of $476$\,nHz (about 12\,nHz, or 50 m\,s$^{-1}$ faster than the mean rotation rate for that latitude and depth). This 50 m\,s$^{-1}$ prograde speed seems to track the emerging region quite well during the 37.5 days, i.e., the time period in which we extract the data. \figref{fig:bvslices} shows the snapshots of the extracted ${\bf B}$ and ${\bf v}$ fields at the time instance shown in Figure 10 of \citet{Fan+Fang:2014}. Note that the extracted $v_{\phi}$ is relative to the reference frame co-rotating at the tracked rotation rate of $476$\,nHz.

The strong azimuthal field (negative $B_{\phi}$) in the flux emergence region at the center of the slice has a peak field strength of  about $9800$ G, and rises with a velocity ($v_r$) that is not significantly different from the upflow speed in other convective giant cells. The emerging region of strong azimuthal field shows a tilt, with the leading end (leading in the direction of rotation) being closer to the equator than the following. \figref{fig:emg3d} shows a 3D view of the magnetic flux bundle that gives rise to the emerging region. \figref{fig:emg3d}(a)(b)(c) show a set of 3D field lines of the emerging flux bundle colored with $B_{\phi}$, and (pink) isosurfaces of strong downflows with $v_r$ reaching -110 m\,s$^{-1}$, viewed from 3 different perspectives. \figref{fig:emg3d}(d) shows the same perspective as \figref{fig:emg3d}(c) but with the spherical surface of $B_{\phi}$ at the depth of the extracted slice (see \figref{fig:bvslices}(c)) included to show the location of the field lines in relation to the emerging flux region. As discussed in \citet{Fan+Fang:2014}, the emerging flux bundle is sheared (by the local prograde moving giant-cell flow) into a ``hairpin'' shape, with the leading ``hook'' of the hairpin structure adjacent to a strong downflow (the pink isosurface just in front of the hook in Figures \ref{fig:emg3d}(a)(b)(c)). This type of arrangement is found to cause the earlier formation of the stronger and more coherent leading sunspots as will be discussed in \sectref{sec:asym}. This flux bundle will be the focus of the this study and is referred as the \emph{central flux bundle} hereafter.

In addition to the central flux bundle, there is another significant flux bundle at $-8$ deg latitude (hereafter, referred as the \emph{southern flux bundle}) that has emerged earlier, as can be seen in \figref{fig:bvslices}(c). Because this flux bundle is in the southern hemisphere, the direction of the azimuthal field, as well as the leading and following polarity, is opposite to that of the central flux bundle (in the northern hemisphere), and is consistent with the Hale's polarity law. Moreover, the southern flux bundle is also embedded in a giant convective cell and rises with a velocity similar to the upflow speed of giant cells at this depth. In particular, at the beginning of the extracted time series, it is already passing through the extracted layer, while the central flux bundle has not broken into this layer. Consequently, the southern flux bundle is the first flux bundle that reaches the photosphere, as shown in \sectref{sec:res_photo}

\subsection{Flux Emergence Simulations}\label{sec:setup_fem}
To study the rise of the flux bundles generated in the convective dynamo to the solar surface, we conduct a series of numerical simulations of the upper convection zone of the Sun with the MURaM code \citep{Voegler+al:2005,Rempel+al:2009,Rempel:2014}. The MURaM code solves the fully compressible MHD equations and accounts for the radiative transfer in grey or non-grey solar atmosphere and a realistic equation of state. This treatment allows the code to reproduce the granulation and radiation as observed at the surface of the Sun, as well as solar convection from the scale of granules to supergranules. 

The code has been extensively used to study magnetoconvetion inside sunspots \citep{Schuessler+Voegler:2006}, penumbral structures and dynamics \citep{Rempel+al:2009.sci,Rempel+al:2009}, quiet Sun magnetism \citep{Danilovic+al:2010,Rempel:2014}, emergence of magnetic flux \citep{Cheung+al:2007,Cheung+al:2008} and formation of active regions \citep{Cheung+al:2010,Rempel+Cheung:2014}.	

The setup of flux emergence simulations in the present paper is similar to that in \citet{Rempel+Cheung:2014} in many aspects. The simulation domains are cartesian and equidistant in the horizontal dimensions. We use the grey-atmosphere assumption for the radiative transfer in simulations presented in this paper. The simulations\footnote{excluding the high resolution case $98\times8$\_hres that is restarted from a snapshot of the $98\times8$ run at about $t{=}23$\,h. The high resolution run is only design to study fine structures and dynamics on much shorter time scale than the life time of a sunspot.} start from a thermally relaxed convection simulation in which radiative cooling at the photosphere (corresponding to the solar energy flux) is in balance with convective energy transport throughout the subphotospheric part of the domain. The convection zone contains a mixed polarity magnetic field that is maintained by a small-scale dynamo, as shown in \citet{Rempel:2014}. 

The lateral boundaries are periodic for all variables. The top boundary of the simulation domain is located 640\,km above the average optical depth unity ($\tau{=}1$) level of the quiet Sun. Vertical flows at the top boundary are strongly damped. The other hydrodynamic variables are set to be symmetry with respect to the top boundary. The magnetic field is a potential field calculated from the vertical magnetic field in the upper most cell in the simulation domain.  

The essential difference is that we use the data extracted from the solar convective dynamo simulation by \citet{Fan+Fang:2014} as a time-dependent bottom boundary. It is important to note that the purpose of this setup is \emph{not} taking these simulated data as a ground truth that exactly reproduces what happens in the Sun. Our main aim is to use a more realistic flux emergence setup in terms of dynamically consistent velocity and magnetic field components. We implement these extracted data as the bottom boundary driver for a series of simulations with various domain width and depth, so that we can test many factors that may impact the emergence of magnetic flux and the formation of active regions in the photosphere. The full list of numerical experiments is presented in \tabref{tab:list}.
 
 \begin{table*}
\center
\caption{Summary of numerical experiments \label{tab:list}}
\begin{tabular}{l|cccc|cccc|c}
\hline
\hline
Run Name & Width\tablenotemark{a} & Depth & Mesh & Grid Spacing & & Factors & &~& Refer to \\
~ & [Mm] & [Mm] & [$N_x^2\times N_z$] & [km$^3$] & $v_{z}$ ($f_{vz}$) & $t$ ($f_t$) & $L_{\rm cs}$($f_L$) & Flux\tablenotemark{b}($f_{\Phi}$) & Figures. \\
\hline
~\,98$\times$8 & ~\,98.304 & ~\,8.192 & ~\,512$^2 \times$128 & 192$^2 \times$64 & 5.0 & 1/5 & 0.08 & 1.0 & \ref{fig:btau1}--\ref{fig:emerg_time}, \ref{fig:asymm}--\ref{fig:vzbz} \\
~\,98$\times$8\_hres & ~\,98.304 & ~\,8.192 & 2048$^2 \times$512 & ~\,48$^2 \times$16 & 5.0 & 1/5 & 0.08 & 1.0 & \ref{fig:highres} \\
~\,98$\times$18\_$t1$ & ~\,98.304 & 18.432 & ~\,512$^2 \times$288 & 192$^2 \times$64 & 5.0 & 1/5 & 0.08 & 1.0 &  \ref{fig:emerg_time} \\
196$\times$8 &196.608 & ~\,8.192 & 1024$^2 \times$128 & 192$^2 \times$64 & 5.0 & 1/5 & 0.16 & 2.0 & \ref{fig:emerg_time} \\
~\,98$\times$18 & ~\,98.304 & 18.432 & ~\,512$^2 \times$288 & 192$^2 \times$64 & 5.0 & 2/5 & 0.08 & 2.0 & \ref{fig:emerg_time}, \ref{fig:vsurf}, \ref{fig:vsurf2}, \ref{fig:btau1_multi}, \ref{fig:bvert_multi} \\
196$\times$18 &196.608 & 18.432 & 1024$^2 \times$288 & 192$^2 \times$64 & 5.0 & 2/5 & 0.16 & 4.0 & \ref{fig:emerg_time}, \ref{fig:btau1_multi}, \ref{fig:bvert_multi} \\
196$\times$32 &196.608 & 32.768 & 1024$^2 \times$512 & 192$^2 \times$64 & 5.0 & 2/5 & 0.16 & 4.0 & \ref{fig:emerg_time}, \ref{fig:flow32}, \ref{fig:btau1_multi}, \ref{fig:bvert_multi} \\
\hline
\end{tabular}
\tablenotetext{a}{Equidistance in horizontal dimensions}
\tablenotetext{b}{Relative to the 98$\times$8 case, as defined by \equref{equ:flux_fac}}
\end{table*}
 
The data extracted from the dynamo simulation contain magnetic structures in both the northern and southern hemispheres, and the magnetic structures have already developed properties in their corresponding hemispheres in the rotating convective shell, e.g., the opposite leading polarities. When these magnetic structures are transported into the domain of the MURaM simulations, it allows us to study the robustness of the near-surface layer evolution for emerging flux bundles with properties of both hemispheres simultaneously. In the rest of the paper we will still use the terms: \emph {equator}, \emph{northern} and \emph{southern hemispheres} , to refer to the corresponding locations in the original convective dynamo simulation that is done in a rotating spherical shell.

The rotation effect is omitted in the current MURaM simulations that cover up to about 15\% of the depth of the whole convection zone. However, even the largest domain depth considered in these simulations is still relatively small compared to the depth of the convective dynamo simulation, which is about 90\% of the depth of the whole convection zone. Therefore, we expect that the influence of the rotation in the upper most layer of the convection zone on the properties of the emerging flux bundles that have been developed in the dynamo simulation would be rather insignificant. Neither do we take into account the effect of the near-surface shear layer, where the rotation rate has a strong radial gradient, which may also affect the emergence of the flux bundle through the upper most layer of the convection zone. The aim of the this paper is mainly to investigate how the magnetic structures generated in solar convective dynamo give rise to sunspots in the solar surface, and if their properties would be retained by the photospheric sunspots. The effects of the rotation and near-surface shear would be very interesting for further investigation in future works, but are beyond the scope of this study.

\subsection{Coupling of the two simulations}\label{sec:setup_couple}

As described in \sectref{sec:setup_ff}, we select a large region centered at an emerging flux bundle at 30\,Mm below the photosphere, and extract the data with a cadence of 3 hours for the first 300 hours, and 6 hours for the rest of the time. The extracted variables are the three components of the velocity field, i.e., $v_{r}$, $v_{\theta}$, and $v_{\phi}$, and horizontal components of the magnetic field, i.e., $B_\theta$ and $B_\phi$. The flux emergence simulations are done in a Cartesian domain, and we directly use the $\phi$, $\theta$, and $r$ components in the spherical coordinate as the $x$, $y$\footnote{$v_{y}{=}-v_{\theta}$ and $B_{y}{=}-B_{\theta}$, because positive direction of $\theta$-axis is southward.}, and $z$ components in the Cartesian coordinate, respectively. This gives us 
$$
\mathbf{U}_{\rm dynamo}^{\rm ori} = \left[v_x^{\rm ori}, v_y^{\rm ori}, v_z^{\rm ori}, B_x^{\rm ori}, B_y^{\rm ori}\right].
$$
Generally speaking, the coupling strategy is using the data from the dynamo simulation to set the velocity and magnetic field in the ghost cells at the bottom of the flux emergence simulation, while density and plasma energy are set by the typical boundary condition for magneto-convection simulations of MURaM (described in details in \sectref{sec:setup_gc} ). Note that we only need to prescribe $B_x$ and $B_y$ at the boundary, then the vertical gradient of $B_z$ is constrained by the solenoidal nature of $\mathbf{B}$, and hence $B_z$ at the bottom boundary is evaluated by this gradient and  $B_z$ in the computational domain. However, we can not simple copy $\mathbf{U}_{\rm dynamo}^{\rm ori}$ into the ghost cells at the bottom boundary. The details of implementation are described in the following paragraphs.

\subsubsection{Adapting to the periodic lateral boundaries}
\begin{figure*}
\includegraphics[width=0.96\textwidth]{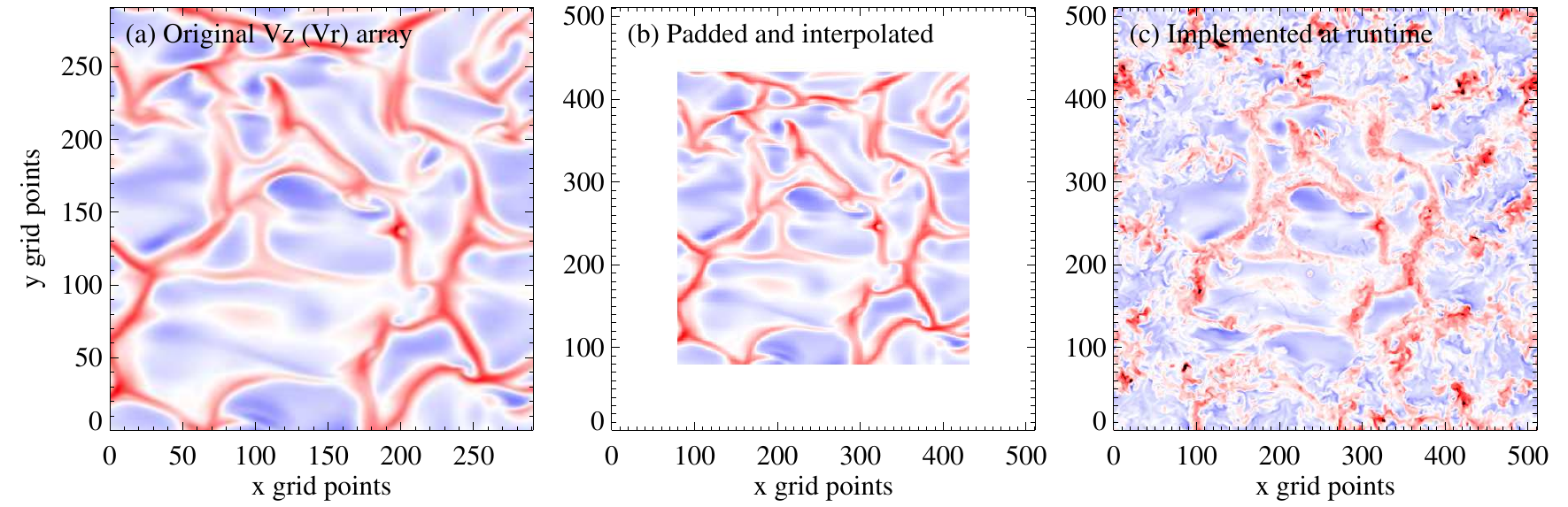}
\caption{An example of the implementation of the bottom boundary. The simulation case shown here is the $98\times18$ run (see \tabref{tab:list}). (a) The original $292\times292$ array of $v_{z}$ (i.e., $v_r$ in the data extracted from the flux emergence simulation). (b) The original $292\times292$ array is padded with a rim of 66 grid points and the padded array is interpolated to the $512\times512$ mesh of the flux emergence simulation. (c): During the simulation the bottom boundary is set by a linear combination of the data from the dynamo simulation (in the central part) and the typical boundary condition for magneto-convection simulations (in the outer rim). See \sectref{sec:setup_gc} for details. \label{fig:bot_bound}} 
\end{figure*}

The data extracted from the dynamo simulation is cropped from the spherical surface, thus it is non-periodic. We must modify these data to make them compatible with the periodic lateral boundaries of the flux emergence simulation. On each side of the $292\times292$ array (i.e., one 2D array for each variable) we attach a padding of 66 grid points with values of zero. Then the padded array of $424\times424$ grid points is interpolated to fit the horizontal mesh of the flux emergence simulation (e.g., a 512$\times$512 array or 1024$\times$1024 array). Hereafter $\mathbf{U}_{\rm dynamo}^{\rm ori}$ refers to the variable arrays that have been padded and interpolated. \figref{fig:bot_bound} (a) and (b) give an example of the padding and interpolation of the original array. Eventually the central part of the domain is set by the data from the dynamo simulation, and the padded outer rim is set by the typical boundary condition for magneto-convection simulations with MURaM, as described in \sectref{sec:setup_gc}.

\subsubsection{Rescaling the data extracted from the dynamo simulation}\label{sec:setup_scaling}
The original horizontal extension of $\mathbf{U}_{\rm dynamo}^{\rm ori}$ is about 800\,Mm (i.e., 70$^{\circ}$ in the $\theta$ and $\phi$ directions). The flux emergence simulation in the upper most layer of the convection zone requires a sufficiently high resolution to properly resolve the granules, and hence the simulations would become very computationally demanding, if we were to keep the original horizontal spatial scale. The most practical approach is to downscale the horizontal extension of the original data to 100\,Mm -- 200\,Mm, which are much more affordable in computational resources, and can still well represent a typical active region.

In summary, we conduct numerical experiments with 98.304\,Mm and 196.608\,Mm wide domains (hereafter 98\,Mm and 196\,Mm, respectively) and with a set of different depths (see \sectref{sec:res_cases} for more details). As mentioned above, we add a padding to the original data array and the padded data array is interpolated to match the mesh and domain size of the flux emergence simulation. The corresponding scaling factor in length, $f_{\rm L}$, is 0.08 for runs with a 98\,Mm wide domain, and 0.16 for those with a 196\,Mm wide domain.

In fact the spatial scale of the convection and magnetic structures in the dynamo simulation might have been overestimated for the following 3 reasons: (1) The data is extracted near the closed top boundary of the dynamo  simulation. Horizontally diverging upflows might artificially expand the emerging flux region. (2) The higher diffusivity in the numerical experiment can affect the spatial scales of convection and magnetic structures. \citep{Jones+Kuzanyan:2009} showed that the size of  convective cells may scale with $E_{k}^{1/3}$, where $E_{k}$ is the Ekman number that evaluates the ratio of viscous forces to Coriolis forces. (3) Global 3D simulations likely overestimate the convective Rossby number of the convection zone \citep[e.g.,][]{Featherstone+Hindman:2016}, which leads to enhanced convective velocity power on large scales. 

Therefore, if the solar convective dynamo simulation could have been done with parameters that are closer to the realistic values, we would expect it to yield smaller convective cells, which is inline with the trend of the rescaling we implement in this paper. 

Otherwise, if one would naively take the original spacial scale and magnetic field strength in the data extracted from the dynamo simulation, the amount of flux that emerges to the photosphere, as well as the sizes of active regions that form through the flux emergence, would be very unrealistic. Therefore, the downscaling in the spatial scale does provide a model setup that is more comparable with the situation on the real Sun.

The magnetic flux needs to be accompanied  by a proper upflow, so that the flux can be transported into the simulation domain. Previous simulations used to implement a uniform upflow of 500\,m\,s$^{-1}$\citep[see the setup in][]{Rempel+Cheung:2014}, which is likely too fast. A recent study by \citet{Birch+al:2016} showed that when the upflow imposed at the bottom boundary of a simulation is significantly faster than the mean convective upflow velocity, the rising magnetic structure will drive a large-scale diverging flow pattern at the surface that is inconsistent with the observed flows. Based on comparisons between a series of numerical simulations down to 20\,Mm below the solar surface and observations, they suggested that the upper limit of the rising speed of magnetic flux at 20\,Mm depth is about the local convective upflow velocity. Although this conclusion is drawn at 20\,Mm depth, this work still provides a sensible guideline for numerical experiments with other domain depths.

On the other hand, we found in earlier experiments that the flux emergence speed has to be comparable to the typical convective upflow for flux emergence to happen. Therefore, the most reasonable setup for this study is rescaling the mean vertical velocity in the extracted data to match that in the MURaM simulations. For all simulations in the present paper, the vertical velocity imposed at the bottom boundary is $v_z{=}v_{z}^{\rm ori}\cdot f_{vz}$, where $v_{z}^{\rm ori}$ is the original vertical velocity extracted in the dynamo simulation and $f_{vz}{=}5.0$. This value is chosen, so that the mean of $v_z$ is about the same as the mean of vertical velocity at 18\,Mm depth in MURaM simulations. We also explore other values for $f_{vz}$, such as $f_{vz}{=}3.0$ and $f_{vz}{=}10.0$. It is clear that the former can not provide effective magnetic flux transport into the domain, while the latter induces unrealistic flow patterns similar to earlier simulations. The values of other variables (i.e., $v_x$, $v_y$, $B_x$, and $B_y$) remain unchanged.

For the production runs in this paper, we do not enforce that $\nabla\cdot(\rho u)$ remains zero when the vertical velocity is rescaled. However, we also carry out a series of numerical experiments where we rescale the horizontal velocities according to $f_{vz}$, $f_{L}$, and the density scale height at the depth of the bottom boundary, such that $\nabla\cdot(\rho u)$ remains the same as that of the data from the dynamo simulation. By comparing the the results from simulations with and without the enforcement, we found the difference to be insignificant. First of all, this is because the vertical velocity imposed at the bottom boundary plays a more important role in the evolution of the flow field and magnetic field than the horizontal flow velocity, as discussed later in this paper. The horizontal flow field near the bottom of the domain evolves mostly in response to the evolution of the vertical flow field. Secondly, the rescaling factor that would be required to keep the $\nabla\cdot(\rho u)$ unchanged is actually relatively close to unity. For all the production runs considered in this study, that factor is never larger than two.

Finally, the time ($t$) of $\mathbf{U}_{\rm dynamo}^{\rm ori}$ is scaled by $t{=}t^{\rm ori}\cdot f_{t}$, where $t^{\rm ori}$ is the original time, and the speed up factor, $f_{t}$, is $1/5$ for most of the simulations. This is motivated by that faster emergence should last for a shorter time and the basic reference is $f_{t}f_{vz}{=}1$, so that rescaling in vertical velocity and time does not change the amount of magnetic flux transported through the boundary. However, we notice that $f_{t}{=}2/5$ is vitally necessary for simulations with deeper domain (i.e., 18 or 32\,Mm). When the domain is deeper, a significant amount of magnetic flux injected from the bottom boundary will not reach the photosphere but is turned over by the convective downflows. Therefore, using a slightly larger $f_{t}$ helps to provide more magnetic flux input, so that a sufficient amount of the magnetic flux can reach the photosphere. In addition to that, the convective pattern in the deeper layers has a longer life time, which means using a $f_{t}$ larger than that for 8\,Mm deep run would not be unrealistic.

As we use combinations of different domain sizes and rescaling parameters, the amount of magnetic flux advected into the domain can be different. The flux transported by an emerging flux bundle moving through the bottom boundary can be evaluated by 
\begin{equation}\label{equ:flux}
\Phi = \int_{\Delta t} B_{\rm h} L_{\rm cs} v_{\rm up} {\rm d}t,
\end{equation}
where $B_{\rm h}$ is the horizontal magnetic field strength, $L_{\rm cs}$ the length of intersection between the cross section of the horizontal flux tube and the bottom boundary, $v_{up}$ the upflow velocity, and $\Delta t$ the time it takes the flux bundle to move through. As described in \sectref{sec:setup_couple}, scaling factors (i.e. $f_{L}$, $f_{t}$, and $f_{vz}$) have been applied on the time and vertical velocity of $\mathbf{U}_{\rm dynamo}^{\rm ori}$. Consequently the factor of the magnetic flux transported through the bottom boundary relative to the reference 98$\times$8 run is given by 
\begin{equation}\label{equ:flux_fac}
f_{\Phi}=\frac{f_{vz} f_{L} f_{t}}{\left(f_{vz} f_{L} f_{t}\right)_{98\times8}},
\end{equation}
where the denominator denotes the combination of the rescaling parameters for the $98\times8$ run. Please refer to the section "Factors" in \tabref{tab:list} for a list of the scaling factors.

\subsubsection{Implementation in the ghost cells}\label{sec:setup_gc}
In solar magneto-convection simulations with the MURaM code, the mass flux and magnetic field are set to be symmetric with respect to the bottom boundary. We refer to the ghost cell values defined in this way as $\mathbf{U}_{\rm MURaM}$. Finally, actual ghost cells ($\mathbf{U}_{\rm gc}$) are set by a combination $\mathbf{U}_{\rm MURaM}$ and $\mathbf{U}_{\rm dynamo}$ (which refers to $\mathbf{U}_{\rm dynamo}^{\rm ori}$ after the padding, interpolation and rescaling described earlier), as defined in \equref{equ:combine} below.

We first use a linear interpolation in time to obtain $\mathbf{U}_{\rm dynamo}$ at each timestep, and the ghost cell values are determined by 
\begin{equation}\label{equ:combine}
\mathbf{U}_{\rm gc} = \mathbf{U}_{\rm dynamo}~M + \mathbf{U}_{\rm MURaM} (1-M),
\end{equation}
and $M$ is a 2D array defined as
\begin{equation}
M_{i,j}=
\begin{cases}
1 & \text{ if } r<0.24\\ 
3.4-10r & \text{ if }  0.24 < r < 0.34\\ 
0 & \text{ if } r>0.34,
\end{cases}
\end{equation}
where $r$ is the normalized distance to the center of the domain. Therefore, the central part of the domain is set by $\mathbf{U}_{\rm dynamo}$. The outer rim, where $\mathbf{U}_{\rm dynamo}$ is padded with zero values, is filled with ordinary magneto-convections at this depth. Between these two regimes there is a linear transition with a width of 10\% of the domain width. \figref{fig:bot_bound} (c) shows the actual $v_z$ implemented at the bottom boundary in the $98\times18$ run (i.e., corresponding to $\mathbf{U}_{\rm gc}$). In particular, at this depth, the convective patterns in the data from the dynamo simulation (after rescaling) match very well with the intrinsic convective patterns in MURaM simulations which are considered as a good representation of those in real solar convection zone.

After setting the velocity and magnetic field in the ghost cells at the bottom boundary, the entropy of upflows is set to a fixed value, which is found to lead to the solar energy flux (at the surface) under quiet sun conditions. The mean pressure is extrapolated into the ghost cells as described by \citet{Rempel:2014}, while pressure fluctuations are damped. Finally, the density is derived from the pressure and entropy via the equation of state.  Thus, the thermal quantities at the bottom boundary are completely determined by the near-surface layer flux emergence simulation with the MURaM code, without using any input from the dynamo simulation. This is essentially required by reproducing the correct solar energy flux at the surface, as well as the correct stratification, which are crucial for realistic simulations.

\subsection{Summary of numerical experiments}\label{sec:res_cases}

We conduct numerical experiments with different domain sizes and rescaling parameters to explore the emergence of magnetic flux bundles through the upper most convection zone. In the analysis of the results, we focus on the simulation with a 98.304\,Mm wide and 8.192\,Mm (hereafter 8\,Mm for simplicity) deep domain. The top boundary is about 640\,km above the photosphere. The density contrast between the bottom and the photosphere is about 2000 out of the total density contrast of 10$^{6}$ in the convection zone. The contrast through the entire box would be even larger, because of the fast dropping density above the photosphere. 

We also carry out simulations with deeper domain depths of 18.432\,Mm and 32.768\,Mm (hereafter 18\,Mm and 32\,Mm, respectively), and with the horizontal width extending to 196\,Mm, to explore the influences of greater density contrasts and larger-scale convective motions. As a result, the bottom-to-photosphere density contrasts in the 18\,Mm and 32\,Mm runs are increased to about 12000 and 40000, respectively. Given that the convective dynamo simulation already included a density contrast of the order of 100, the coupled model is able to cover the density contrast of the whole convection zone. For completeness of the model setup we also carry out one simulation with a 196\,Mm wide and 8\,Mm deep domain. A detailed list of the simulations considered in this paper is summarized in \tabref{tab:list}.

For most of the production runs, we use a horizontal grid spacing of 192\,km and a vertical one of 64\,km. With the numerical scheme implemented in the MURaM code \citep{Rempel:2014}, this resolution is sufficient to produce granulation and radiation flux that are in good agreement with the real Sun. To study the fine structure and dynamics in the penumbra, we use a horizontal grid spacing of 48\,km and a vertical one of 16\,km, which are necessary to properly resolve filamentary structures in sunspot penumbrae. In particular, the vertical resolution is more crucial because the driver of the Evershed flow is concentrated in a very thin layer near the $\tau{=}1$ surface\citep{Rempel:2011}. The high resolution simulation is only evolved for a period of about 2 hours, because the computational demand for full evolution would be too large.

\begin{figure*}
\center
\includegraphics{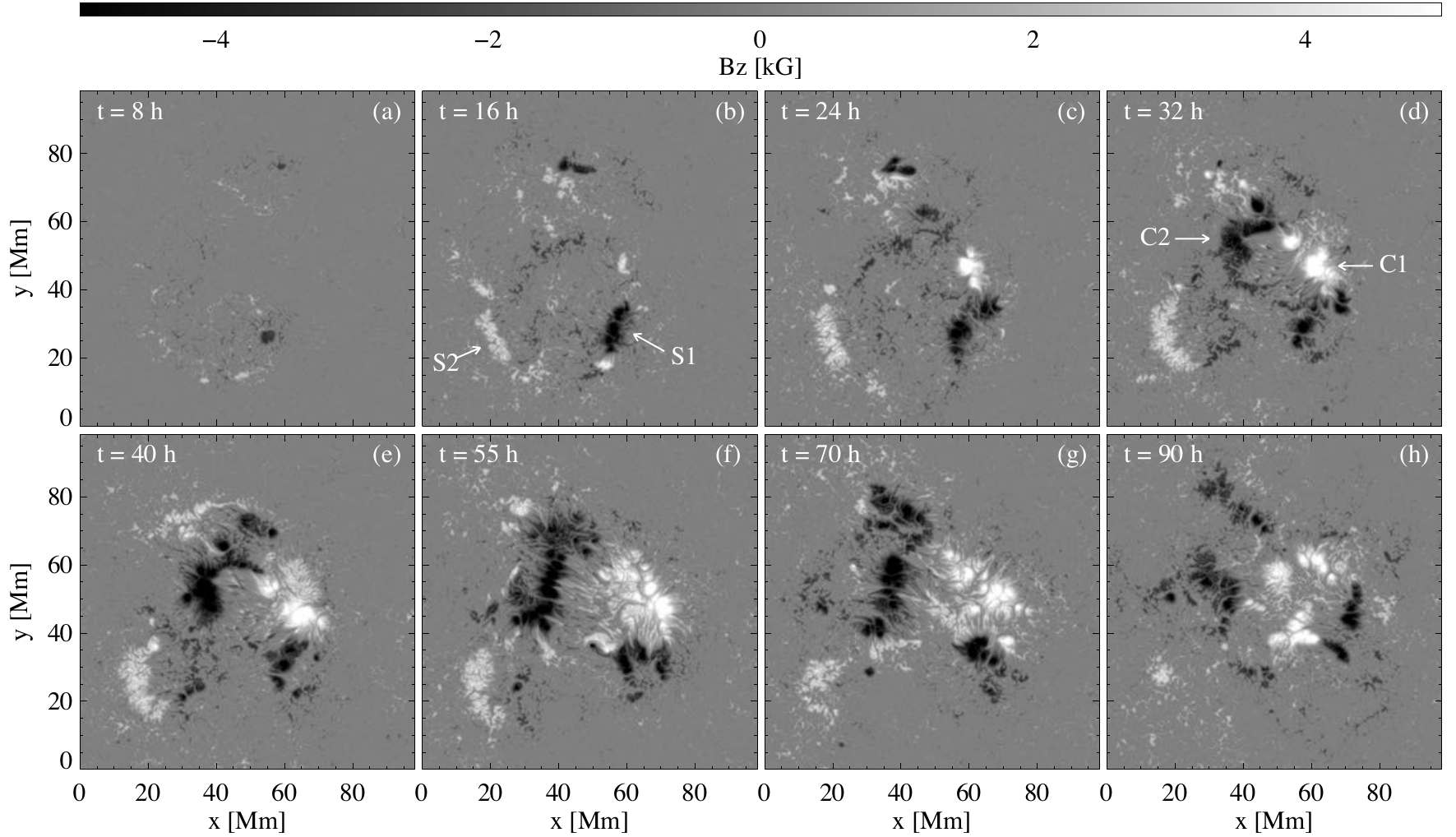}
\caption{Vertical magnetic field at the $\tau{=}1$ layer showing the evolution of active regions in about 4 days. This can be compared with magnetograms from observations. An animation is available in the online version of the paper. \label{fig:btau1}}
\end{figure*}

\section{Emergence of Magnetic Flux Bundles and Formation of Sunspots and Active Regions}\label{sec:res}

\subsection{Magnetic flux emergence at the photosphere}\label{sec:res_photo}

\begin{figure*}
\center
\includegraphics{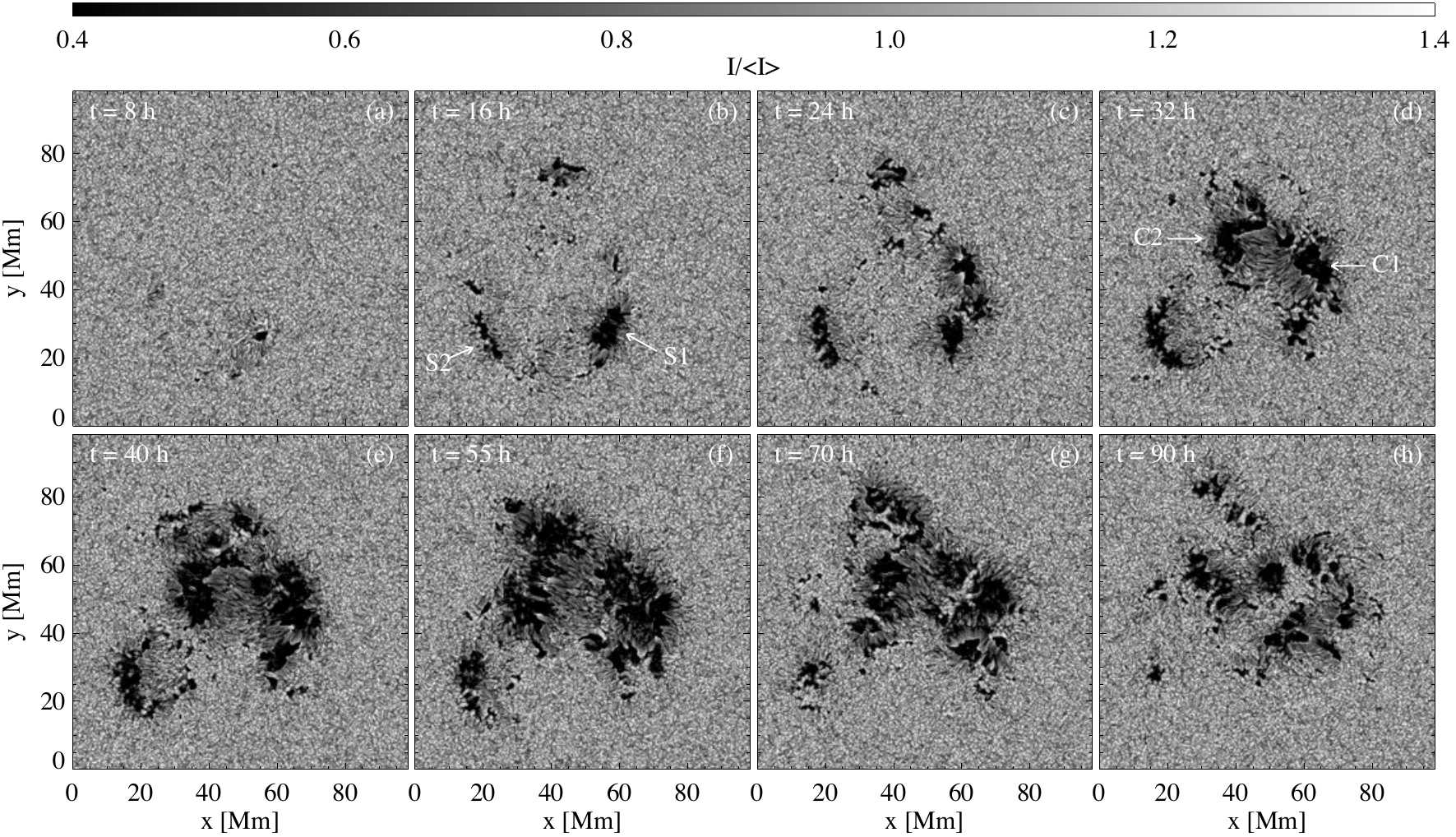}
\caption{Continuum intensity images for the same time period as in \figref{fig:btau1}. The intensity is normalized by the mean intensity in the quiet Sun. An animation is available in the online version of the paper. \label{fig:iout}}
\end{figure*}

\begin{figure}[t!]
\center
\includegraphics{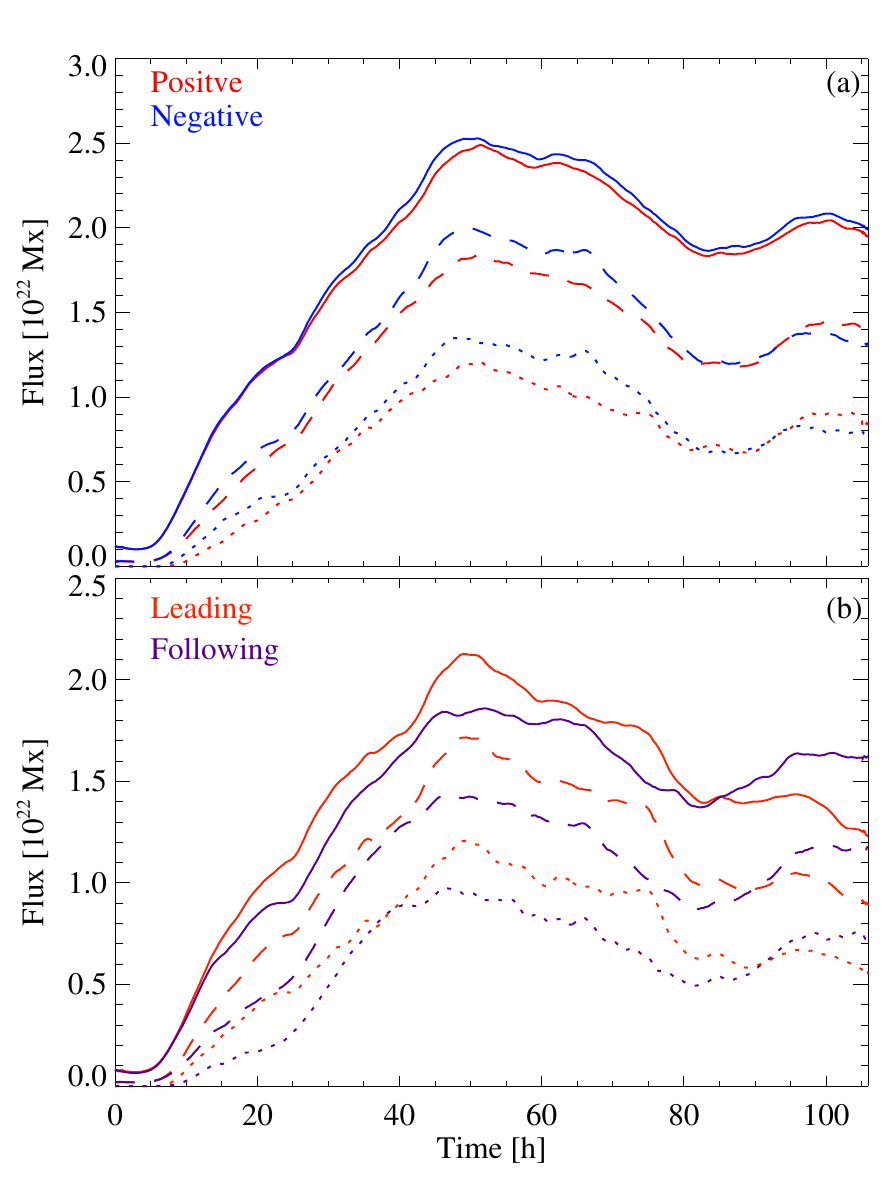}
\caption{Magnetic flux at the $\tau{=}1$ layer as a function of time in the $98\times8$ run. (a): Red lines are for the positive polarity and blue the negative. Solid lines are integrated over the entire photosphere, dashed lines are integrated within regions where the continuum intensity ($I$) is less than 0.8\,${<}I{>}$, and dotted lines are integrated within regions where $I$ is less than 0.5\,${<}I{>}$, where ${<}I{>}$ is the mean intensity in the quiet Sun. (b): Evolution of the magnetic flux in the region only covering the two major sunspot pairs. Red lines represent the leading polarity, which is the positive in the original northern hemisphere in the convective dynamo simulation ($y{>}$37.8\,Mm when mapped to the flux emergence simulation) and negative in the original southern hemisphere in the convective dynamo simulation ($y{<}$37.8\,Mm). Blue lines are for the following polarity. Solid, dashed, and dotted lines are defined in the same way as in the upper panel.  \label{fig:mag_flux}}
\end{figure}

The magnetic flux bundles are advected through the bottom boundary by upflows at an average speed of about 150\,m\,s$^{-1}$. The magnetic flux further rises through the upper most layers of the convection zone and breaks into the photosphere. \figref{fig:btau1} shows the vertical magnetic field at the $\tau{=}1$ layer in 8 snapshots that cover a period of about 4 days. \figref{fig:iout} shows the corresponding images of continuum intensity normalized by the mean intensity in the quiet Sun area. The time stamps are the same as in \figref{fig:btau1}. These two figures allow a direct comparison with magnetrograms and white light images from observations.

As shown in \figref{fig:btau1}(a)(b) most of the magnetic features in the photosphere in the early stage of flux emergence are small-scale magnetic elements of a similar size of solar granules. This reflects that the coherent magnetic flux bundles are broken into smaller pieces by the turbulent convection when it rises through the convection zone. This behavior is consistent with previous flux emergence simulations accounting for the interaction of convection with rising magnetic structures \citep{Cheung+al:2010,Stein+al:2011,Rempel+Cheung:2014}. 

Large-scale magnetic flux concentrations start to form from coalescence of small-scale magnetic elements. At $t{=}8$\,h (\figref{fig:btau1}(a)) we find a small but evident concentration of magnetic flux in the negative polarity at ($x{\approx}60$\,Mm, $y{\approx}25$\,Mm), which is surrounded by small-scale magnetic elements of the same polarity. In the intensity image it can be clearly seen as a small dark pore. To the left of the area dominated by the negative polarity field, there is some magnetic field in the positive polarity distributed along an arch from ($x{\approx}20$\,Mm, $y{\approx}30$\,Mm) to ($x{\approx}60$\,Mm, $y{\approx}$10\,Mm). By $t{=}16$\,h (\figref{fig:btau1}(b)) both structures have developed into sunspots with kilo G strong magnetic field. The negative polarity spot appears in the intensity image as a dark spot with a diameter of more than 10\,Mm, while the positive polarity spot seems to be smaller and follows the shape of its magnetic structure. This sunspot pair corresponds to the southern flux bundle in the original convective dynamo simulation. Hereafter, we refer to the leading spot as \emph {Spot S1} and the following as \emph{Spot S2}, as indicated by arrows in \figref{fig:btau1}(b) and \figref{fig:iout}(b).

In the time period between $t{=}16$\,h and $t{=}32$\,h another sunspot pair forms in a similar manner in the central part of the domain.  As indicated by arrows in \figref{fig:btau1}(d) and \figref{fig:iout}(d), we refer to the leading and following spots as \emph{Spot C1} and \emph{Spot C2}, respectively. It is clear that this sunspot pair originates from the central flux bundle. Since the central flux bundle locates in the original northern hemisphere of the convective dynamo simulation, the leading polarity of this sunspot pair (positive) is opposite to that of the sunspot pair formed by the southern flux bundle (negative). The position of the original equator in the convective dynamo simulation is at $y{=}37.8$\,Mm for all simulations with 98\,Mm wide domain. 

\figref{fig:mag_flux} (a) shows the magnetic flux integrated over the entire photosphere as a function of time. It also shows the flux in areas of less than 80\% mean intensity of the quiet Sun (${<}I{>}$), which generally corresponds to the areas occupied by sunspots, and the flux in areas of less than 50\% ${<}I{>}$, which represent the dark umbra areas of sunspots. Hereafter, we refer to them as flux in the sunspots and flux in the umbrae, respectively. By $t{=}32$\,h about 1.8$\times10^{22}$\,Mx magnetic flux has emerged into the photosphere, and given rise to two major sunspot pairs and many scattered smaller magnetic structures. We can see that the magnetic flux in the photosphere keeps increasing until $t{\approx}45$\,h. The flux in the sunspots and that in the umbrae of sunspots also increases in the same trend, which suggest that the sunspots get continuously intensified by newly emerged magnetic flux. 

However, the process of decay and fragmentation takes over around $t{=}50$\,h. In the rest of the evolution the total magnetic flux and the fluxes in the sunspots do not decrease significantly. This implies that the process is mostly destruction of magnetic features. It can also be seen from \figref{fig:btau1} and \figref{fig:iout} that the large sunspots gradually lose their coherence and break into smaller magnetic structures. The fragmented pieces still maintain a relatively strong field strength and hence appear as dark features in the intensity images. As the simulation evolves longer, we anticipate that the sunspots will eventually decay. However, that would be beyond the scope of this study.

\subsection{Subsurface evolution of emerging flux bundles}\label{sec:res_vert}
\begin{figure*}[t!]
\center
\includegraphics{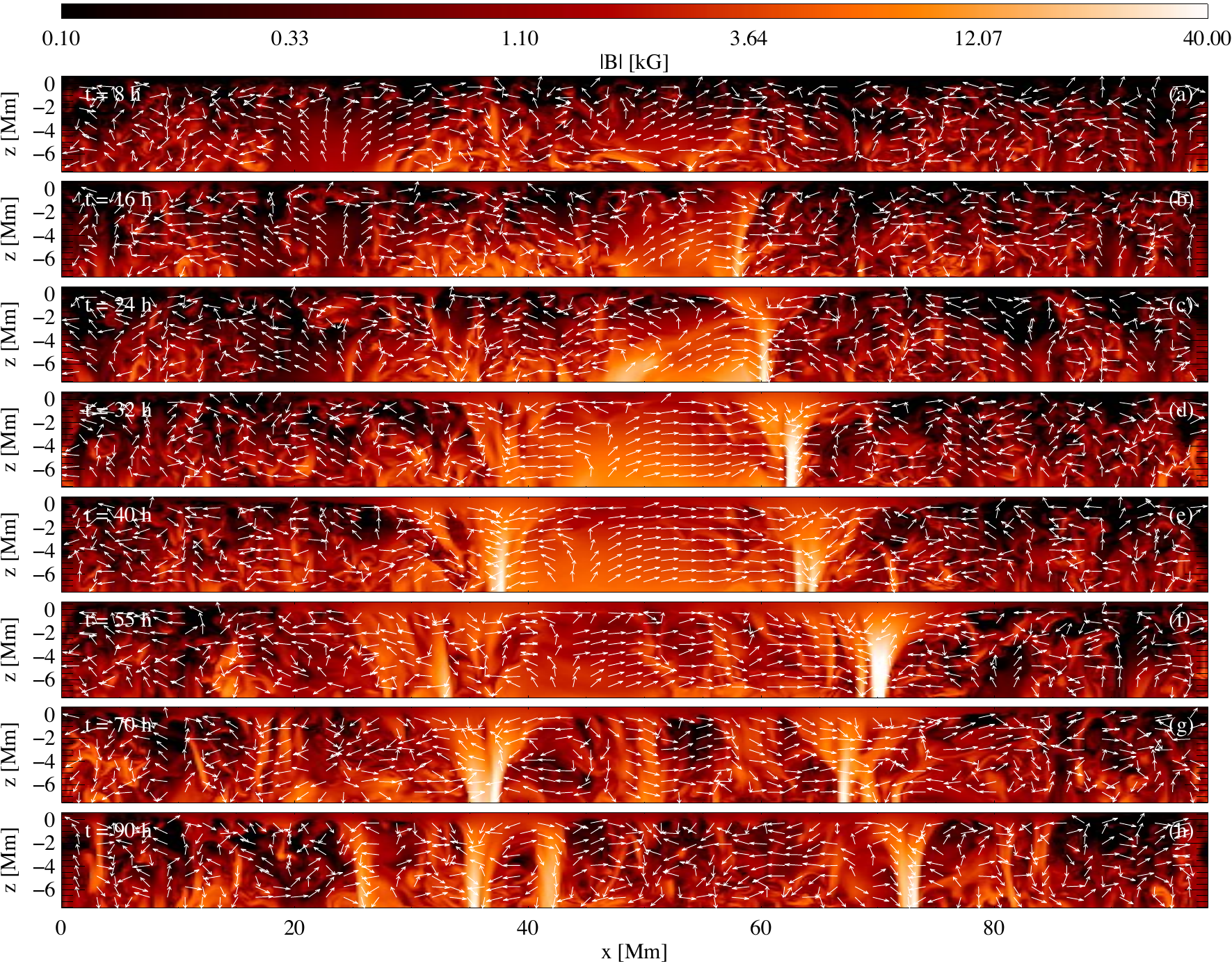}
\caption{Magnetic field strength ($\left | \mathbf{B} \right |$) in a vertical slices ($x-z$ plane) through the center of the domain ($y{=}49$\,Mm in the 98$\times$8 run). Time stamps are identical to those in \figref{fig:btau1}. Arrows show the normalized $v_x$ and $v_z$ in the $x-z$ slice. Note that the length of the arrows doest not correspond the speed of the flow. \label{fig:bvert}}
\end{figure*}

\figref{fig:bvert} shows the evolution of the magnetic field strength in the vertical slice (i.e., in the $x{-}z$ plane)  through the center of the domain. The times stamps are identical to those in \figref{fig:btau1} and \figref{fig:iout}. It generally traces the rise of the central flux bundle. However, in a complex simulation such as the one shown here, it is not possible to define a single vertical slice that exactly cuts through both sunspots and captures their entire evolution. Nonetheless, \figref{fig:bvert} can still reveal the flow pattern and magnetic structure beneath the sunspot pair, which is not directly observable. Throughout the paper we define the mean height of the $\tau{=}1$ surface as $z{=}0$\,Mm.

In the early stage when the center flux bundle has not emerged through the bottom boundary, the domain is dominated by the magnetic field that is maintained by the small-scale dynamo (see $t{=}8$\,h in \figref{fig:bvert}). A few hours later (see $t{=}16$ and 24\,h) the emerging flux bundle rises to $z{\approx-}2$\,Mm as a relatively coherent structure. However, from this depth to the photosphere, the coherent flux bundle starts to fragment into smaller magnetic elements that eventually reach the photosphere as shown in \figref{fig:btau1}. 

While the (predominantly horizontal) flux bundle cannot reach the photosphere as a coherent structure, we do find that a coherent vertical magnetic flux concentration forms in the convection zone. This flux concentration has a rather small and consistent width below $z{\approx}-6$\,Mm, but dramatically expands in the upper most 2\,Mm. In the photosphere this magnetic structure corresponds to Spot C1 (or the pore before Spot C1 is well formed). Therefore, the sunspot in the simulation presented here is a monolithic magnetic structure that can reach deep in the convection zone (see more discussions in \sectref{sec:res_deep}). Hereafter, we use the term sunspot to refer to not only the strong magnetic flux concentration that is seen as a dark spot in the photosphere, but also the corresponding magnetic structure in the convection zone. The evolution in the early stage also suggests that the sunspot beneath the photosphere may have already well formed before the strong flux concentration forms in the photosphere, which means the sunspot may form from bottom to top.

\figref{fig:bvert}(e) shows that at $t{=}40$\,h both Spot C1 and Spot C2 have formed. The convection zone between the two sunspots (i.e., between $x{\approx}35$\,Mm and $x{\approx}65$\,Mm) is still filled with the emerging flux from the central flux bundle. At $t{=}55$\,h we find that the magnetic field strength in the region between the two sunspots is clearly reduced, which suggests that the bulk of the central flux bundle has completely emerged through the convection zone. This timing is in line with \figref{fig:mag_flux}(a) that shows the photospheric flux peaks at about $t{=}50$\,h. Moreover, at this time stamp, the two sunspots also start to fragment into multiple vertical flux concentrations, which is consistent with the behavior seen in the photosphere (\figref{fig:btau1}). Sometimes we don't see the sunspots in the later stage in \figref{fig:bvert}. This does not mean that the sunspots have decayed, but because, once the sunspots start to fragment, the vertical slice actually quite often cuts through the lanes between the fragmented magnetic structures. Therefore, \figref{fig:bvert} must be examined together with \figref{fig:btau1}.

\subsection{On the emerging speed of magnetic flux bundles}\label{sec:res_temer}
\begin{figure*}
\center
\includegraphics{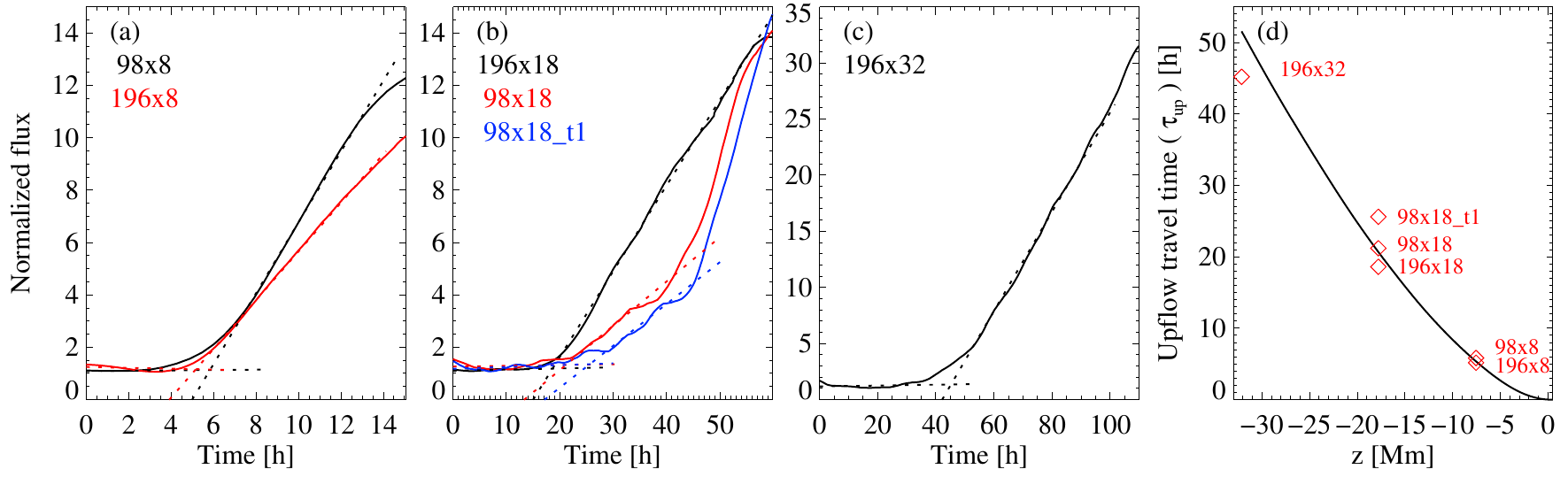}
\caption{Normalized magnetic flux in the photosphere as functions of time. See definitions in \sectref{sec:res_temer}. (a): Simulations with 8\,Mm deep domain. (b): Simulations with 18\,Mm deep domain. (c) Simulation with 32\,Mm deep domain. See list of simulation setup in \tabref{tab:list}. Dashed lines show linear fitting to the pre-emergence stage and emergence stage for each run. (d): Convective upflow travel time ($\tau_{\rm up}$), which is defined in \equref{equ:upflow}, as a function of depth. The actual travel times of emerging flux tube in 6 runs are indicated by red diamonds.
 \label{fig:emerg_time}}
\end{figure*}

\begin{table*}[t!]
\center
\caption{Comparison between $\tau_{\rm travel}$ and $\tau_{\rm up}$}\label{tab:emer}
\begin{tabular}{r|cccccc}
\hline
\hline
Run Name                  & 98$\times$8 & 196$\times$8 & 98$\times$18 & 98$\times$18\_t1 & 196$\times$18 & 196$\times$32 \\
\hline
$\tau_{\rm travel}$ [h] & 5.83             & 5.17                & 21.2               & 25.6                      & 18.6                  &  45.2                 \\
$\tau_{\rm up}$ [h]      & 5.34             & 5.37                & 20.7               & 20.7                      & 20.7                  & 51.6                 \\   
\hline
\end{tabular}
\end{table*}

For the analysis of the emerging speed and the travel time of the flux bundles in the convection zone, we focus on the sunspot pair formed at about $y{\approx}30$\,Mm in \figref{fig:btau1} and \figref{fig:iout}(a)(b)(c), which is corresponding to the southern flux bundle in the convective dynamo simulation. Note that this flux bundle is already emerging into the domain at $t{=}0$, therefore, the travel time ($\tau_{\rm travel}$) of the flux bundle through convection zone can be directly evaluated by the time instant when Spot S1 and Spot S2 start to appear.

\figref{fig:emerg_time}(a)(b)(c) show the unsigned magnetic flux in the photosphere in the area corresponding to the original southern hemisphere in the convective dynamo simulation (i.e., $y{<}37.8$\,Mm for domain width of 98\,Mm, and $y{<}75.6$\,Mm for domain width of 196\,Mm). These magnetic flux curves have been normalized by their minimum values, respectively, and grouped by the depth of the corresponding simulation cases.

In all six cases selected for this analysis, we see a common feature that the magnetic flux in the photosphere remains unchanged in the very early stage (e.g. before $t{\approx}5$\,h in runs with 8\,Mm depth, hereafter, pre-emergence stage), and then starts to increase with a rather constant slope (hereafter, emergence stage). The former illustrates the saturated small-scale dynamo, and the latter is a clear signature of the magnetic flux bundle arriving at the photosphere.

To estimate $\tau_{\rm travel}$, we use two linear functions to fit the pre-emergence stage and the emergence stage, respectively. These linear functions are also plotted as dashed lines in \figref{fig:emerg_time}, and their colors are the same as their corresponding magnetic flux curves. Then the coordinate of intersection of these two linear functions in $t$ is considered as $\tau_{\rm travel}$. We note that the $98\times18$ and $98\times18\_t1$ runs have a second emergence stage with a larger slope after $t{\approx}40$\,h. One linear function would not properly fit the two emergence stages simultaneously. Therefore, as we see that the curve between $t{\approx}20$\,h and $t{\approx}40$\,h is already substantially different from that in the pre-emergence stage, we only apply the linear function to the first emergence stage, i.e. the one between $t{\approx}20$\,h and $t{\approx}40$\,h.

In simulations with realistic convection, convective upflows were found to play an important role in transporting magnetic flux, in particular when the field strength is not strong enough to induce a significant buoyancy \citep{Stein+al:2011}. To evaluate the importance of convection in transporting magnetic flux in the simulation in this study, we also estimate travel time of a magnetic element that is solely transported by convective upflows by
\begin{equation}\label{equ:upflow}
\tau_{\mathrm{up}}(z) = \int^{0}_{z^{\prime}{=}z}\frac{\mathrm dz^{\prime}}{\bar{v}_{\mathrm{up}}(z^{\prime})},
\end{equation}
where $\bar{v}_{\mathrm{up}}$ is the mean upflow speed at a given depth $z$. The convective upflow travel time, $\tau_{\mathrm{up}}$, represents the travel time from a certain depth to the surface under the condition that a magnetic element is passively advected through the convection zone by convective upflows, and not impacted by downflows.

We obtain $\tau_{\mathrm{up}}$ as a function of depth using the first snapshot (i.e., the initial condition) of the $196\times32$ run. During this phase, there is no large-scale magnetic field, but only mixed-polarity field that is maintained by a saturated small-scale dynamo in the convection zone. The solid line in \figref{fig:emerg_time}(d) shows $\tau_{\mathrm{up}}$, with $\tau_{\rm travel}$ overlaid as red diamonds. In general, the actual travel time of flux bundle is very well aligned with the convective upflow travel time. This clearly suggests that the emergence of the magnetic flux in these simulations is dominated by the upflow in the convection zone, and hence the rising speed of the flux bundles can be well represented by the mean convective upflow speed at the corresponding depth.

\subsection{Large-scale near-surface flows during flux emergence}
\begin{figure*}[t!]
\center
\includegraphics{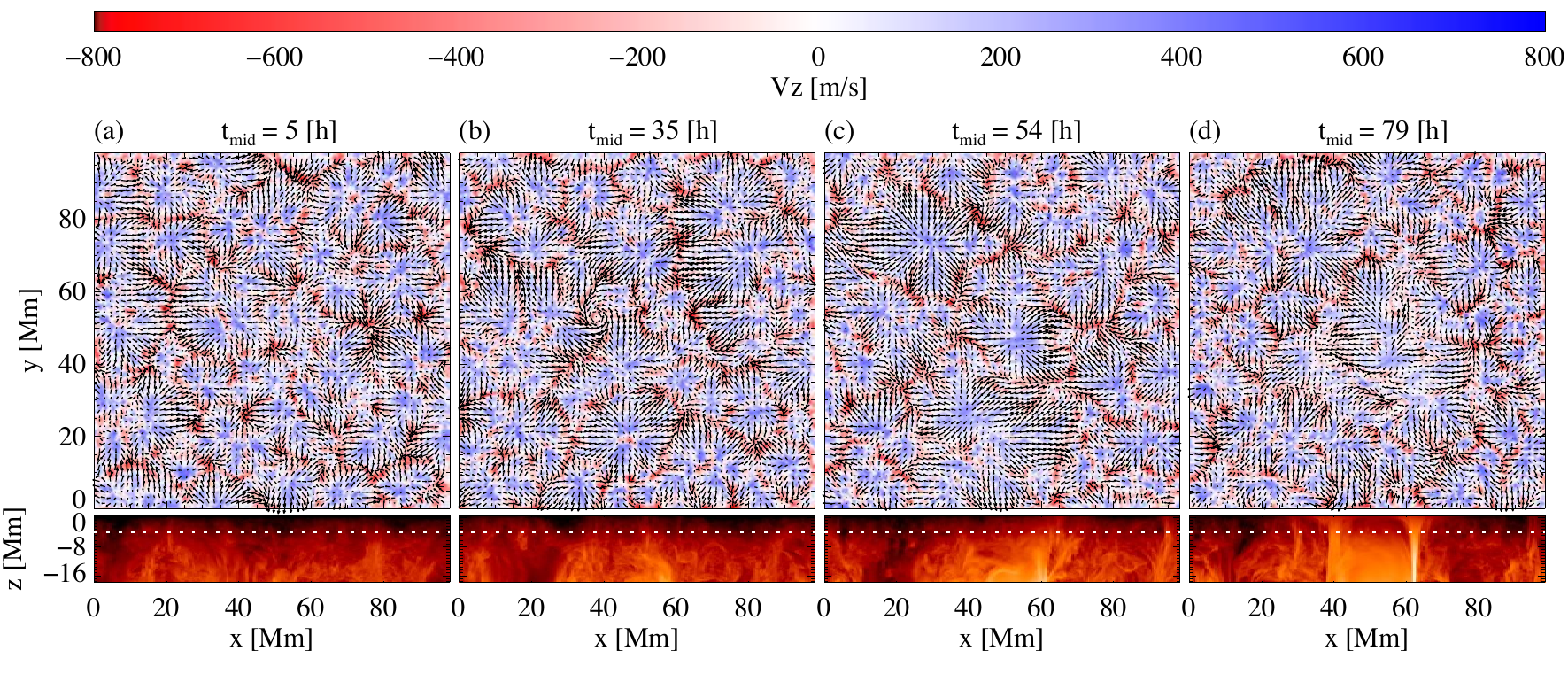}
\caption{Near surface flow patterns before and during flux emergence. The flow and magnetic fields in the 98$\times$18 run are averaged over a time period of 20000 iterations corresponding to about 10 hours. $t_{\rm mid}$ is the middle of each of the averaged time windows. Upper panels plot the vertical velocity in a horizontal slice at 4\,Mm depth below the photosphere (blue for upflows and red for downflows). Arrows show the horizontal velocity. Lower panels show the magnetic field strength in a vertical slice through the center of the simulation domain (i.e., a $x-z$ slice at $y=49$\,Mm). The quantity is shown in the same color scale as in \figref{fig:bvert}. The dotted lines in the lower panels indicate the corresponding depth of the horizontal slice in the upper panels.
\label{fig:vsurf}}
\end{figure*}

\begin{figure}[t!]
\center
\includegraphics{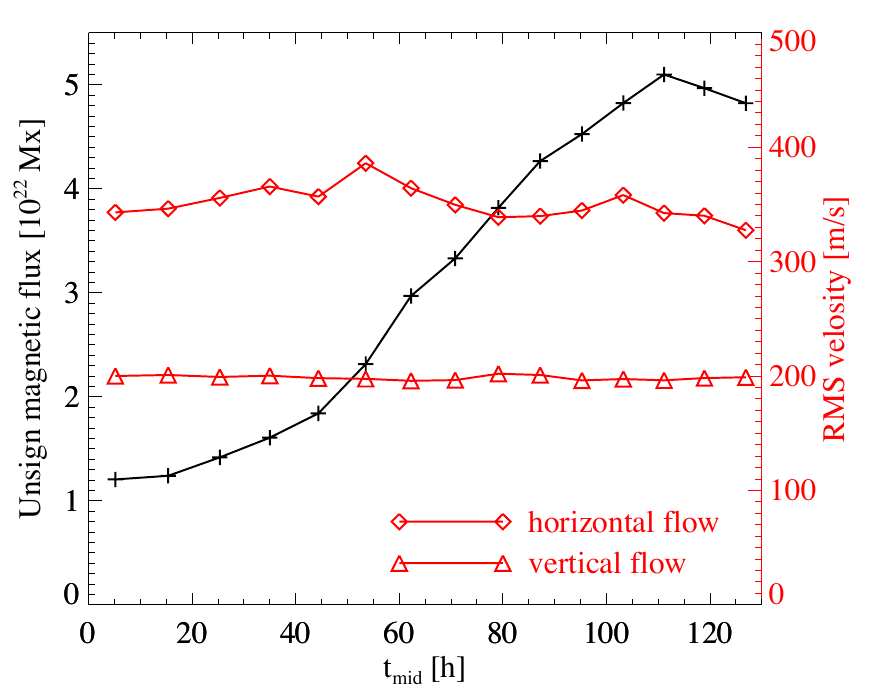}
\caption{Variation of the magnetic flux and flow field at a depth 4\,Mm during the flux emergence. The data are the same in \figref{fig:vsurf}. Black solid line shows the magnetic flux before and during the flux emergence stage. Red diamonds and triangles show the root mean square (RMS) of the horizontal and vertical velocities, respectively.
\label{fig:vsurf2}}
\end{figure}

The flow field in the upper most layer of the convection zone can be inferred by helioseismology. This information gives people a valuable insight on the processes of the flux emergence. \citet{Birch+al:2016} analyzed the horizontal flows near the solar surface during flux emergence events and found the patterns of horizontal flows are not affected by the rising magnetic flux. They also suggested that surface flows in numerical simulations are similar to observed flows only if the rising speed of magnetic structures is comparable to or lower than the convective velocities.

The simulations in this paper consider the emergence of magnetic structures that are much more complex than those analyzed in the study of \citet{Birch+al:2016}. Moreover, the amount of magnetic flux introduced into the domain is also significantly larger in our simulations. Therefore, it is very interesting to investigate the near-face flows before and during flux emergence, and compare them with the results of \citet{Birch+al:2016}. 

For this analysis, we use the $98\times18$ run that has a very similar numerical setup to simulations analyzed by \citet{Birch+al:2016}. However, We decide not to perform the same analysis as they did, because helioseismology inversion in active regions with complex strong magnetic features is extremely challenging (Braun, private communication). Here we only focus on a qualitative comparison between the large-scale near-surface flow field in the simulation and that inferred by helioseismology from observations. For this purpose, we calculate the average of the magnetic and flow fields for every 20000 iterations, which corresponds to a time interval of about 10 hours. This temporal average removes information of the small-scale and high frequency granular motions in the near-surface layer, and highlights the long time-scale and large spatial-scale flows.

\figref{fig:vsurf} shows the vertical velocity in a horizontal slice at 4\,Mm depth below the photosphere (upper panels) and the magnetic field strength in a vertical $x-z$ slice through the center of the domain (lower panels). $t_{\rm mid}$ is the middle of each time interval. The four time stamps shown in \figref{fig:vsurf} represent (a): No magnetic flux emerged, (b): The central flux bundle starts to emerge, (c): Spot C1 starts to form, and (d): Spot C1 and Spot C2 are formed in the photosphere, respectively.

The dominant flow pattern in the near-surface layer before the magnetic flux emergence are super-granulation-scale flows as shown in \figref{fig:vsurf}(a). The vertical flow field shows cells of upflow and lanes of downflow. The horizontal flow shows a clear diverging pattern in upflow regions and converging pattern at downdraft lanes. When magnetic flux bundles introduced from the bottom boundary emerge through this layer and give rise to sunspots in the photosphere (\figref{fig:vsurf}(b)--(d)), the general flow pattern in this layer does not seems to change.

Furthermore, we calculate the total unsigned magnetic flux, as well as the mean vertical and horizontal velocities, in the horizontal layer shown in \figref{fig:vsurf} over a time period of about 130 hours . The results are plotted in \figref{fig:vsurf2}. The unsigned magnetic flux significantly increases about 5 times during the flux emergence, and reaches more than $5\times10^{22}$\,Mx at $t{\approx}110$\,h. Meanwhile, the RMS velocities of the horizontal flow only changes very insignificantly, and that of the vertical flow remains almost constant.

The behavior of the horizontal flow in a near-surface layer is also well consistent with the observations by \citet{Birch+al:2016}. This consistency is not entirely surprising, because the mean upflow speed at the bottom boundary is scaled to about the mean convective upflow speed at that depth. This setup has been suggested by \citet{Birch+al:2016} to produce a near-surface flow pattern that is more consistent with observations.

\subsection{Large-scale flow beneath and in the vicinity of the sunspot}\label{sec:res_deep}
\begin{figure*}[t!]
\center
\includegraphics{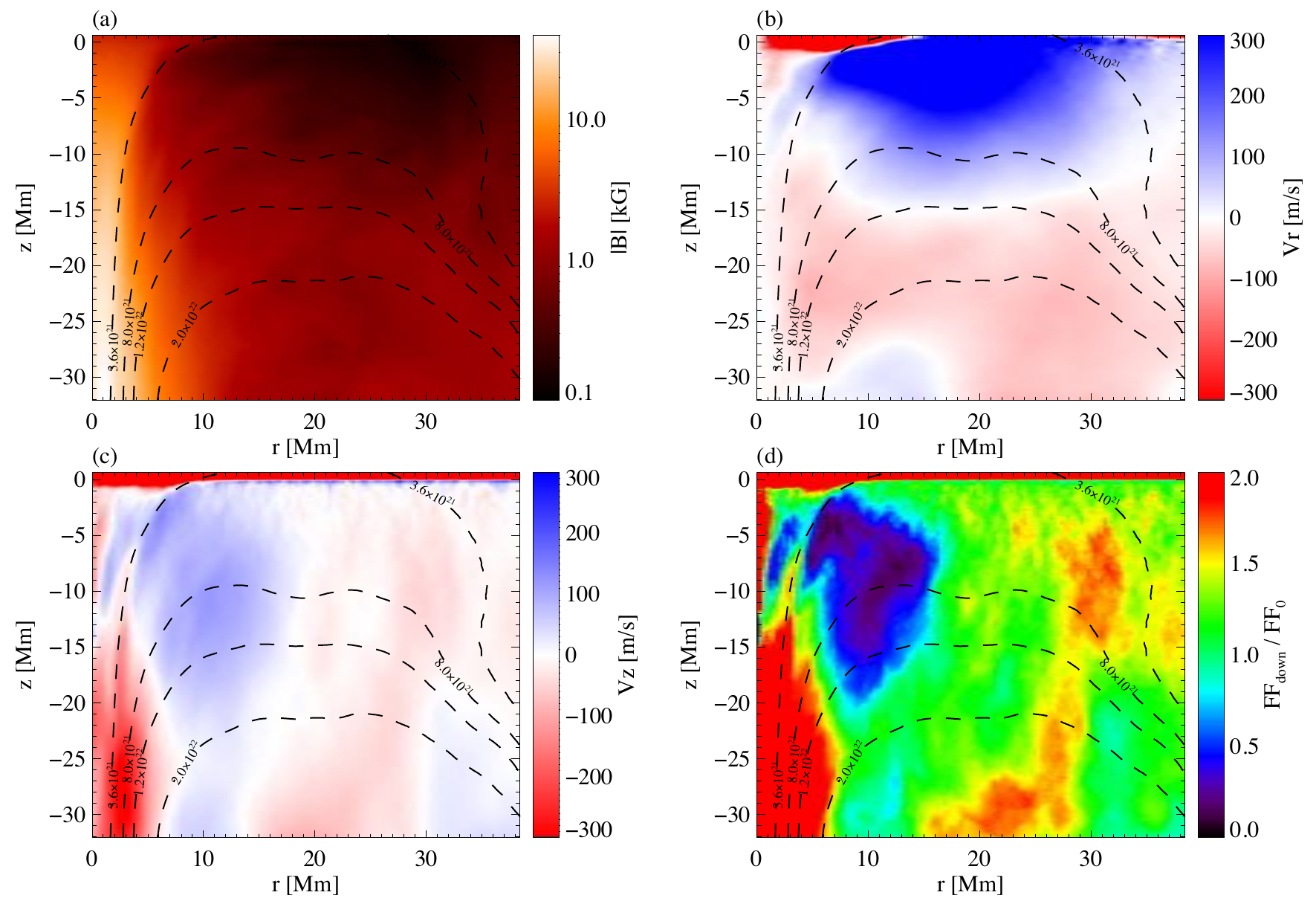}
\caption{Azimutally averaged magnetic field strength and flows beneath Spot C1 and in its vicinity in the $196\times32$ run. (a): The magnetic field strength. Dashed contour lines are for the vertical magnetic flux defined in \equref{equ:phi}. (b): Azimuthally averaged radial velocity. Blue shows outflows (away from the sunspot) and red color indicates inflows. (c): Azimutally averaged vertical velocity. Blue shows upflows and red shows downflows. (d): Azimuthally averaged downflow filling factor relative to the average downflow filling factor in the convection zone at $t{=}0$. \label{fig:flow32}}
\end{figure*}

The magnetic structures and flows beneath sunspots are extremely important for understanding their formation and decay. In particular, numerical simulations that can produce convective motions similar to those of the real Sun, have been used to investigate how the magnetic field structures and flows interact \citep[see reviews by][and reference therein]{Moradi+al:2010}.  Compared with previous simulations that had domain depths of 20\,Mm or less \citep[e.g.,][]{Rempel:2011.sub,Stein+Nordlund:2012}, the $196\times32$ run presented in this paper extends to 32\,Mm below the solar surface, and provides a density contrast that is about three times larger than that in a 20\,Mm deep domain. It also has a large horizontal domain extent that allows development of large-scale convection cells with much less impact from the lateral boundary. These features make the $196\times32$ run an exceptional case to study the magnetic and flow fields beneath a sunspot.

In the following we focus on Spot C1, which is the strongest and most stable sunspot in the simulations. To reduce the impact of temporary fluctuations and better illustrate persistent flows and magnetic structures, we average all quantities in full 3D cubes over a period of 30 hours after Spot C1 has formed in the photosphere. Then we set the center of Spot C1 in the magnetogram at the $\tau{=}1$ layer as the origin of a local cylindrical coordinate and calculate the azimuthal average of the magnetic field strength, radial and vertical velocities over the full depth range of 32\,Mm and a radius range of about 40\,Mm.

\figref{fig:flow32}(a) shows the azimuthally averaged magnetic field strength. The overlaid dashed lines indicate surface of constant vertical magnetic flux ($\Phi$) that is evaluated by
\begin{equation}
\Phi(r,z) = \int_{0}^{r}2\pi r^{\prime}\bar{B}_{z}(r^{\prime},z)dr^{\prime},\label{equ:phi}
\end{equation}
where $\bar{B}_z$ is the azimuthally averaged $B_{z}$. The contour lines shown here are for $\Phi$ equals to 3.6$\times10^{21}$, 8.0$\times10^{21}$, 1.2$\times10^{22}$, 2.0$\times10^{22}$\,Mx, respectively. In particular, $\Phi{=}3.6\times10^{21}$\,Mx approximately matches the amount of magnetic flux contained in the sunspot (i.e., areas where $\vert \mathbf{B} \vert > $1\,kG) at the surface. This line outlines the boundary of the sunspot. The well formed sunspot is a monolithic magnetic structure that extend through the entire vertical domain. It has a very coherent and compact anchor at the depth of the 32\,Mm, which is the bottom boundary of the simulation domain. The sunspot has a radius of less than 2\,Mm at that depth. It almost remains about the same radius in the lower half of the domain, and dramatically expands in the last 10\,Mm beneath the surface. 

The contour lines representing larger amount of flux turn back before reaching the surface. This indicates that the azimuthally averaged vertical magnetic field changes its sign from positive to negative at the turning points.  The turning point for $\Phi{\approx}4\times10^{21}$\,Mx is found at the $z{=}0$\,Mm layer. The magnetic flux contained within $r{=}5$\,Mm at the bottom boundary is already about 5 times larger than the flux that reaches the photosphere.

\subsubsection{Flow inside the sunspot}
A downflow dominates inside the sunspot deeper than $z{\approx}-15$\,Mm. The emerging flux bundle is embedded in a giant convection cell. Consequently, the two ends of the flux bundle are rooted in the downdraft lanes of the convection cell. While the magnetic flux emerges towards the surface and leads to the formation of a sunspot in the photosphere, the roots of the flux bundle remain anchored in the downdraft lane and become the footpoints of the sunspots. Therefore the extended downflow along the deeper part of the sunspot is well consistent with the persistent downdraft lane at which the sunspot anchors.

We find a mixture of up and downflows in the shallower part of the sunspot, which is between $z{\approx}-15$\,Mm and the surface. It is known that convection is still present inside a sunspot. For instance, umbral dots are interpreted as a signature of magnetoconvection inside umbrae \citep{Schuessler+Voegler:2006}. Umbral dots present in this sunspot have probably some contribution to the mixed upflows and downflows. We also find that during the first half of the 30 hours, the flow inside the sunspot is mostly downward. It is likely that right after a strong magnetic flux concentration forms, the enhanced net cooling due to suppression of convective heat transport leads to a drainage of the excessive materials in the sunspot. Convection inside the sunspot may only develop after the new equilibrium is established, and yields the mixture of up and downflows we find in the average over the 30 hours after the sunspot has formed. In addition to that, another non-negligible factor is that the sunspot is not a perfectly axis-symmetric structure and has a rather corrugated boundary. Thus, when computing an azimuthal average as defined above, some flows that are actually outside the spot could have been taken as flows inside the sunspot in the azimuthal average.

\subsubsection{Flow in the vicinity of the sunspot}
Panels (b) and (c) in \figref{fig:flow32} show the azimuthally averaged radial and vertical velocities, respectively. The contour lines are identical to those in panel (a). \figref{fig:flow32}(b) shows that the most dominant flow pattern below $z{\approx}-15$\,Mm outside the sunspot is an inflow that horizontally extends for almost 40\,Mm. Close to the sunspot boundary, which is illustrated by the line of $\Phi{=}3.6\times10^{21}$\,Mm, the inflow speed quickly diminishes, and the flow pattern turns into a downflow, as shown in \figref{fig:flow32}(c) between $r{\approx}5$\,Mm and $r{\approx}$10\,Mm. 

The downflow in this area has the same origin as that inside the sunspot, i.e., it is mostly from the downdraft lane at the leading side of the emerging flux bundle. The flow at this depth is anelastic, thus a downflow is always related with a horizontal converging flow. Given that the downflow here is imposed by the boundary condition, it is clear that the downflow drives the converging flow towards the sunspot, which mostly contributes as an inflow component in an azimuthal average.

The converging flow can be somewhat anisotropic, because the emerging flux bundle is on the left (smaller $x$) of the Spot C1. To investigate the contributions to the inflow from different directions, we also calculate the averages for an azimuthal angle of $\pi$ on the left and right of Spot C1, respectively, . The former can account for the contribution from the flows induced in the flux bundle, while the latter is to illustrate the contribution from flows in the convection zone without a large-scale strong magnetic field. The results (not plotted here) show that the contribution from the flux bundle side dominates within $r{\approx}20$\,Mm, which is about half of the length scale of the center flux bundle, while the contribution from the right side of Spot C1 contributes in a larger distance. As a side note, the prograde motion of the emerging flux bundle does not contribute to the inflow towards Spot C1, because, as we have noted in \sectref{sec:setup_ff}, the data are extracted in a reference frame that moves with the prograde motion of the flux bundle, which is faster than the mean rotation rate (at that depth and latitude). 

The deep inflow was not evident in simulations in previous studies. The smaller domain depths in those simulations might limit the development of the deep inflow, however, a similar inflow pattern is also present in the $98\times18$ and $196\times18$ runs in this paper, which means that the domain depth is not the determinant factor. The most likely reason is that the development of the inflow requires that the sunspot anchors in an area with a persistent downflow. This is essentially the property of the magnetic and flow fields adapted from the convective dynamo simulation, and is not considered in the setup of the other simulations.

A prominent outflow is present above $z{\approx}15$\,Mm outside the sunspot. The outflow is set off from the boundary of the sunspot and extends for more than 20\,Mm in the radial direction. This flow pattern is similar to those found in simulations of stable sunspots \citep{Rempel:2011.sub,Rempel:2015} and sunspot formation through flux emergence \citep{Rempel+Cheung:2014}. This outflow at the vicinity of a sunspot is very likely to represent the moat flow in observations, which is a large-scale outflow around the sunspot in and beneath the photosphere \citep{Gizon+Birch:2005}. The moat flow is suggested to be a consequence of sunspots altering the dynamics of the convection. On the other hand, the moat flow may play an important role on the stability and evolution of sunspots.

To examine the physical cause of the outflow in this simulation, we perform the same analysis as \citet{Rempel:2015} applied in simulations of stable sunspots. The azimuthal average of the downflow filling factor ($\mathrm{FF}_{\rm down}$) is shown in \figref{fig:flow32}(d). The result is shown relative to $\mathrm{FF}_{0}$, which is the average downflow filling factor in the entire convection zone at the beginning of the 196$\times$32 run. $\mathrm{FF}_{0}$ is about 0.38 and represents the balance between the up- and downflows in the convection zone without large-scale strong magnetic field. When a sunspot is present in the convection zone, the balance is clearly disturbed. In \figref{fig:flow32}(d), we find that the downflow filling factor near the sunspot is significantly reduced by a factor of 7, which is similar to the values reported by \citet{Rempel:2015}. The reduction of the downflow filling factor results in a net upflow, and the excess mass flux has to be balanced by a large-scale outflow.

\citet{Rempel:2015} suggested that two effects can reduce the amount of low entropy materials produced at the solar surface near the sunspot. One effect is the reduction of the surface brightness in the penumbral area, which leads to less radiative cooling.  The other effect is the flow in the penumbral area that can efficiently transport the cool materials mostly horizontally, and prevent them from sinking downwards. These effects cause a deficit of low entropy material that feeds into downflows in the vicinity of the sunspot. As a consequence the downflow filling factor in this area is significantly reduced, leading to a net upflow that requires a large-scale outflow for mass flux balance. The net upflow cannot simply vanish in this area, since horizontal pressure balance enforces a stratification similar to that found in other upflows of the surrounding convection zone.

\subsection{Penumbral structures and radial flows in the penumbra}\label{sec:res_pen}
\begin{figure*}[t!]
\center
\includegraphics{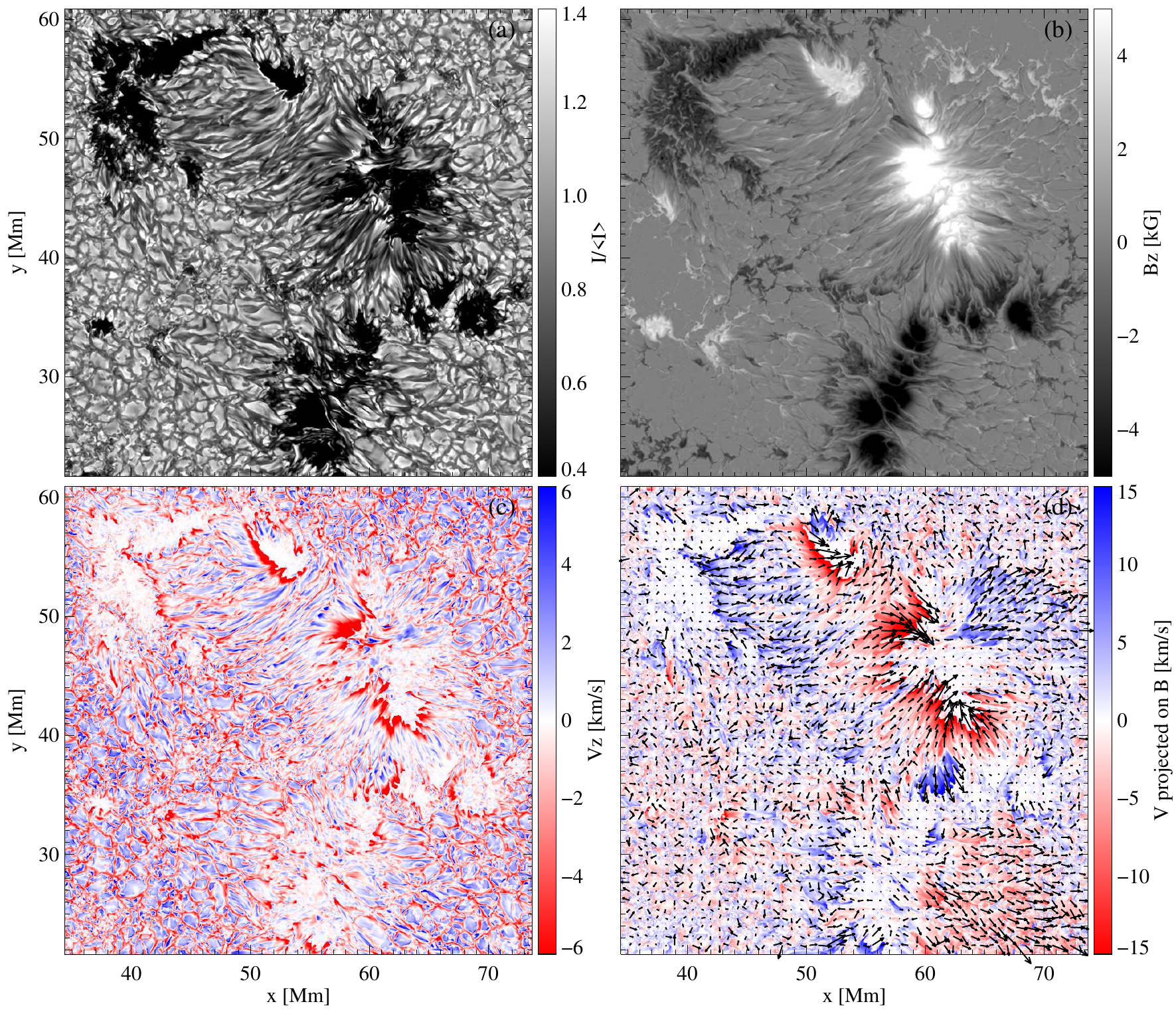}
\caption{Penumbral structures and flows in and near the simulated sunspots in a zoomed view. (a): Continuum intensity. (b): LOS magnetic field in the $\tau{=}1$ layer. (c): LOS velocity in the $\tau{=}1$ layer. (d): Flow along the magnetic field line in the $\tau{=}1$ layer. Arrows show the horizontal flow in the same layer and can help reader to understand the direction of radial flows around in and around the sunspots. An animation is available in the online version of the paper.
\label{fig:highres}}
\end{figure*}

\figref{fig:iout} does show indications of filamentary structures with a brightness of about 70\% ${<}I{>}$ around the dark umbra of the sunspots shortly after their formation starts. These features might correspond to the penumbrae of the sunspots. However, earlier simulations dedicated to study sunspot penumbrae showed that it is crucial to use a sufficiently high resolution. Therefore, we rerun part of the $98\times8$ run with a 4 times higher resolution (see $98\times8$\_hres in \tabref{tab:list}). The rerun is started shortly before $t{=}23$\,h when Spot C1 and Spot C2 have just formed and a smaller flux bundle, which might actually be bifurcated from the main body of the center flux bundle,  is going to break into the photosphere next to the sunspot pair. A zoomed view (focusing on the central part of the domain) of the continuum intensity, vertical magnetic field, vertical velocity and the velocity along the field lines are shown in \figref{fig:highres}.

We find that Spot C1 has the most prominent penumbra, which is comprised of numerous thin and elongated filaments. The brightnesses of the umbra and penumbra are about 24\% and 72\% of the quiet Sun intensity, respectively. As a result of the more realistic-shaped flux bundle used in present simulations, the sunspot umbra has a very irregular shape, and the lengths of penumbral filaments are very anisotropic. This is significantly different from the highly axis-symmetric spots in previous sunspot simulations \citep{Rempel+al:2009,Rempel:2012}. The overall appearance of the sunspot shown in \figref{fig:highres} is quite similar to the "complex" spot shown in \citet{Solanki:2003} and those shown in a recent observation with 0.08$\arcsec$ resolution \citep{Schlichenmaier+al:2016}.

The Evershed effect that is seen as outward flows in the penumbrae of sunspots is a robust observational feature. Previous simulations also successfully reproduced the Evershed effect and explained it as convective upflows being redirected into the radial direction. The analysis of the flows in the radial direction becomes particularly difficult for the sunspots in the present simulation because they have rather corrugated shapes, and many penumbral filaments are actually not along the radial direction. Here we try to use the velocity projected along the magnetic field, $v_{\parallel}$ that is evaluated by $\mathbf{v}\cdot\mathbf{B}/\vert\mathbf{B}\vert$, as a representative of the flows along the penumbral filaments. Since the field lines of a positive polarity spot point outward in the radial direction, a positive/negative $v_{\parallel}$ around the positive polarity sunspot indicates an out-/inflow. \figref{fig:highres}(d) displays a map of $v_{\parallel}$ in the $\tau{=}1$ layer. The arrows show the horizontal velocity (i.e., $v_x$ and $v_y$) in the same layer, which can give a more clear illustration of the flow direction in the penumbra.

Focusing on Spot C1 that has the most prominent penumbral structures, we find fast inflows that can be almost 15\,km\,s$^{-1}$ in the penumbra section facing to the other sunspots (e.g., Spot C2 and Spot S1). The inflow speed increases toward the umbra and peaks at the border between the umbra and penumbra. In contrast, the section facing to the quiet Sun (e.g., the Northwestern part) is dominated by outflows of a few km\,s$^{-1}$, with a much gradual speed variation in the radial direction. In general, the inflow dominates more than 60\% of the penumbra, and only the northwest part shows an extended outflow. We evolve the high resolution run for almost two hours. The aforementioned flow pattern is quite persistent. But it is worthy to note that sometimes outflows can develop in filaments that are originally dominated by inflows. For instance, a filament at ($x{\approx}58$\,Mm, $y{\approx}44$\,Mm) has a clear outflow, which appears as a distinctive blue filament among many red ones in \figref{fig:highres}(d).

The flow in the simulated sunspot penumbra seems to contradict with the ground truth found in observations. The outward Evershed flow is typically observed in penumbrae of well formed sunspots, however, observations of flows in forming penumbrae, in particular when flux emergence is ongoing, are still quiet limited. Given that in this simulation the penumbra filaments with inflows tend to appear in the inner side of active regions (i.e., between sunspots), where the flux emergence is more active, the inflows are probably related to the emerging magnetic flux. When the newly emerged magnetic structures coalesce to the sunspots and contribute to the horizontal magnetic field in penumbra regions, they still need to get rid of the mass load. Because not only is the geometric height of the umbra smaller, but also the pressure in the umbra can be reduced by enhancement of magnetic pressure, the mass load in the newly emerged field lines can drain into the inner penumbra and umbra. In \figref{fig:highres}(c) we can see that the inflows in the penumbra is co-spatial with downflows, which also accelerate toward the umbra and peak at penumbra-umbra boundary. 

The outflows that appear intermittently in some filaments behave very similarly to the normal Evershed flow found in \citet{Rempel+al:2009.sci,Rempel:2012}. As shown in the animation for \figref{fig:highres}, the outflows originate from upflows in the inner end of the penumbral filaments and develop to the outer end by pushing against the inflows. It implies that the driving  mechanism of the normal Evershed flow found in \citet{Rempel:2011} is still in action here. However, the driving force of outflows seems to be overwhelmed by that of inflows in most of the penumbra. Consequently, it leaves an interesting question of whether the flow pattern would change if the simulation is evolved for longer.

The detailed analysis of the formation evolution of the penumbral structures and their dynamics will be addressed in a future paper.

\section{Origin of the Asymmetries in Bipolar Sunspot Pairs}\label{sec:asym}
\begin{figure*}[t!]
\center
\includegraphics{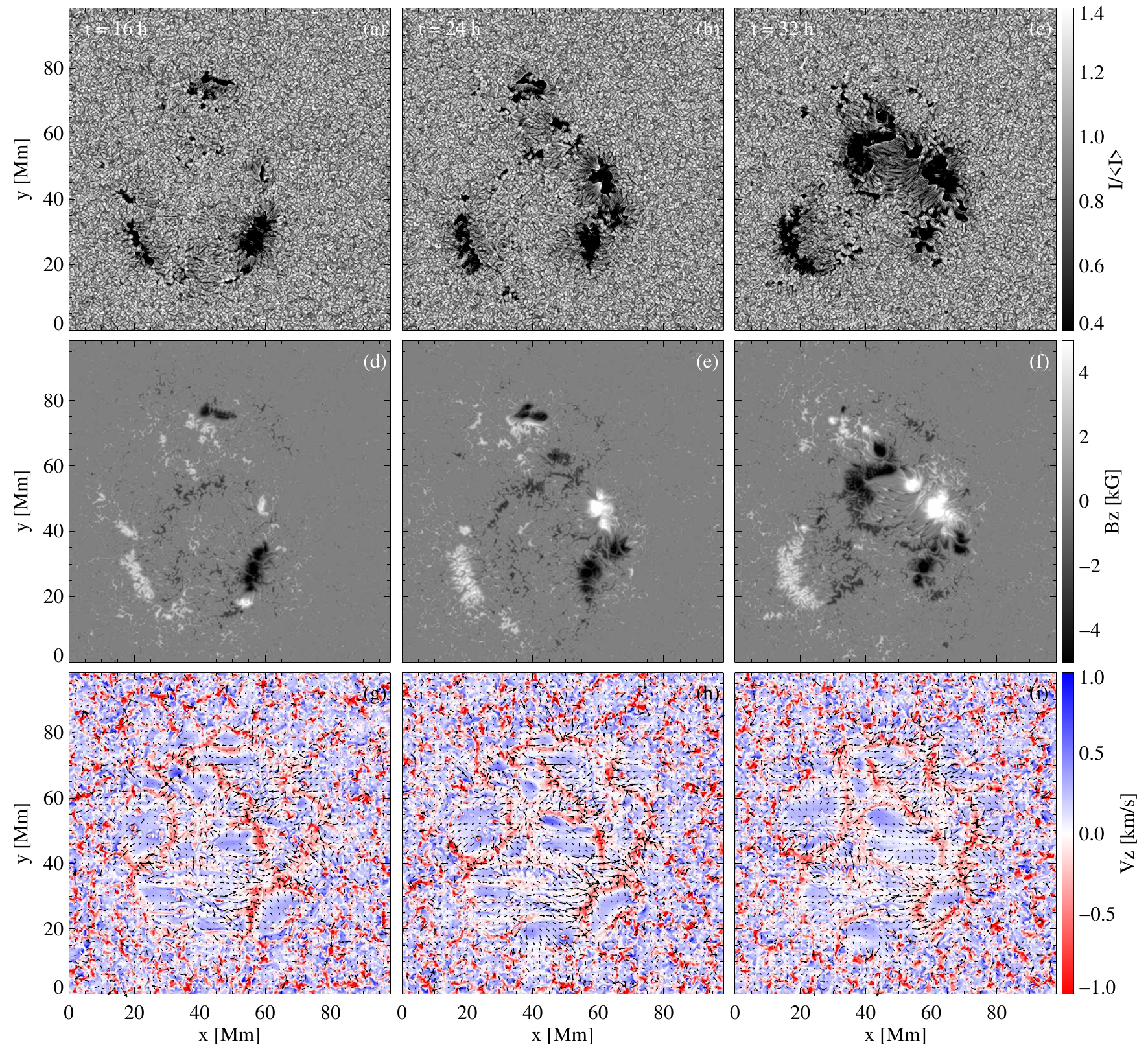}
\caption{Three snapshots corresponding to three stages of the sunspots formation in $98\times8$ run. Uppper panels: Continuum intensity map that can be compared with white light images of the Sun. Middle panels:  Vertical magnetic field at the $\tau{=}1$ layer. Lower panels: Vertical velocity at $z{=}-7$\,Mm. Blue colors represent upflows. Arrows in the lower panels indicate horizontal flows. \label{fig:asymm}}
\end{figure*}

\begin{figure*}[t!]
\center
\includegraphics{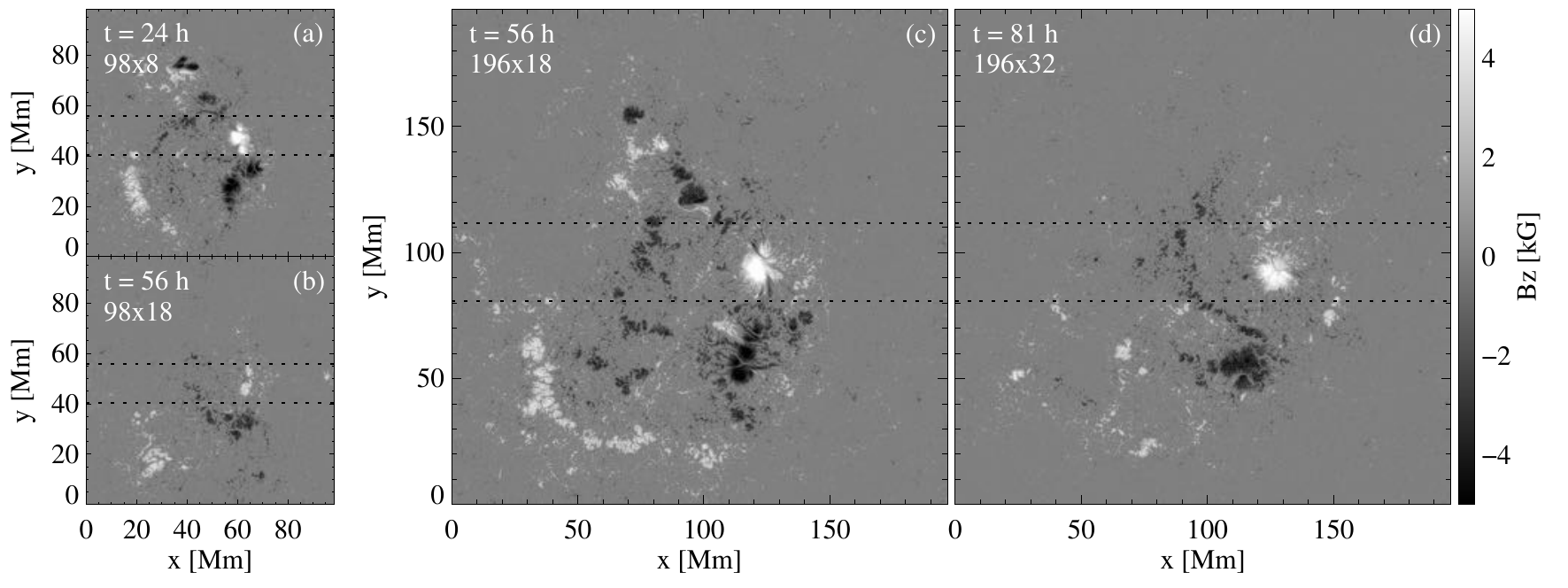}
\caption{Evolution of the vertical magnetic field at the $\tau{=}1$ layer in four simulation cases. The snapshots are chosen to present a similar evolution stage (see \sectref{sec:asym_pho} for details). Dashed lines indicate the regions are used to calculate the average shown in \figref{fig:bvert_multi}. \label{fig:btau1_multi}}
\end{figure*}

The asymmetries between the leading and following sunspots are a fundamental feature of bipolar sunspot pairs. The leading spot in a bipolar sunspot pair is found to be more coherent and of stronger magnetic field than the following spot. Moreover, the leading spots often form earlier than the following spots and have a longer life time \citep[e.g.,][]{McIntosh:1981}, which reveals the asymmetry in their time evolution. The sunspot pairs formed in the photosphere in our simulations reproduce the principle asymmetric properties of real sunspots, which were not reproduced by previous simulations.

In the following, we will demonstrate sunspot asymmetries found in the photosphere of the simulations, and their relation to the asymmetric properties of the magnetic and flow fields in the convection zone, and seek the origin of the asymmetric properties.

\subsection{Asymmetric properties reproduced in the simulations}
\subsubsection{Asymmetric properties of the photospheric sunspots}\label{sec:asym_pho}
\figref{fig:mag_flux}(b) shows the temporal evolution of magnetic flux in a region covering the two major sunspot pairs  (i.e., from $x{=}0$\,Mm to 80\,Mm and $y{=}10$\,Mm to 65\,Mm) in the $98\times8$ run. The magnetic flux is not divided by the positive and negative polarities, but by the leading and following polarities. The dashed and dotted lines present the flux in the sunspots and that in the umbrae, respectively. In this simulation the leading polarity flux accounts for the flux in the positive polarity in the northern hemisphere and that in the negative polarity in the southern hemisphere. Similarly the following polarity flux is the total flux of the negative polarity magnetic field in the northern hemisphere and the positive polarity magnetic field in the southern hemisphere. Note that``{\it northern}"/``{\it southern hemisphere}" in this context refer to regions corresponding to the original northern/southern hemisphere in the convective dynamo simulation by \citet{Fan+Fang:2014}. 

The magnetic flux in the leading polarity is clearly larger than that in the following polarity. The same trend is also valid for the fluxes in the leading and following sunspots, as well as for those in their umbrae.  Because the flux is well balanced in the entire photosphere as shown in \figref{fig:mag_flux}(a), the excessive flux in the leading polarity is possibly balanced either by the opposite polarity flux in the quiet Sun area that is not considered in this estimate or the connecting flux between the two leading spots across the equator. More interestingly, \figref{fig:mag_flux}(b) also shows that the magnetic flux in the leading spots starts to increase earlier. The flux inside the sunspots and that contained in the umbrae, as shown by the dashed and dotted lines, respectively, behaves similarly to the total flux. It suggests that the leading polarity magnetic field has become strong enough to suppress the convection in a considerable area, while the following polarity field has not, i.e., the leading spots form earlier than the following spots. The time lag estimated from \figref{fig:mag_flux} is about a few hours, which is consistent with the estimate one can get from a visual inspection of the evolutions of the vertical magnetic field and continuum intensity that are shown in \figref{fig:btau1} and \figref{fig:iout}, respectively. We note that the flux in the following polarity becomes larger in the late stage simply because part of the fragmented sunspots migrates across the equator and leads to an artificial contribution to the following polarity flux.

To demonstrate how the asymmetry properties develop in the simulation, we focus on Spot C1 and Spot C2, and select three snapshots corresponding to three important stages during the sunspot formation. \figref{fig:asymm} shows the continuum intensity, vertical magnetic field in the photosphere, and the vertical velocity at the $z{=}-7$\,Mm at $t{=}16$\,h, 24\,h, and 32\,h. The first snapshot shown in \figref{fig:asymm}(a)(d)(g) presents the time when the leading spot just appears as a pore without surrounding filamentary structures, and the following polarity field is still too weak to form any evident coherent structure. 

In the second snapshot shown in \figref{fig:asymm}(b)(e)(h), the leading spot has become a coherent and strong sunspot with a diameter of about 10\,Mm. In the continuum intensity image the leading spot exhibits a dark umbra of 24\% of mean intensity in the quiet Sun. Its umbra is surrounded by mostly radial penumbral structures of 72\% of the mean intensity of the quiet Sun. Meanwhile the following polarity field forms a few small pores that are more visible than in the first snapshot. However, the magnetic structures are still very fragmented and the field strength is much lower than that of the leading polarity. 

The third snapshot displayed in \figref{fig:asymm}(c)(f)(i) shows that both the leading and following spots have fully formed. The leading spots has a coherent dark umbra with a diameter of about 15\,Mm, which has grown a little bit since $t{=}24$\,h, and its peak field strength reaches more than 4\,kG. The following spot has grown from a few pores to a significant flux concentration. Its peak field strength also exceeds 4\,kG. However, the following spot is not as coherent as the leading spot.  Penumbral structures are mostly visible between the two sunspots, because this area is dominated by emerging horizontal magnetic field that is likely to facilitate formation of penumbral filaments. In addition, it is clear that the leading spot is closer to the equator, which is consistent with the Joy's law.

Since we have conducted simulations with radically different domain sizes, it is important to examine the robustness of the asymmetries found in the $98\times8$ run. First of all, we take the snapshot at $t{=}24$\,h  in the $98\times8$ run, when the leading spot has formed and the following spot is a group of pores, as the reference. To make a meaningful comparison between different simulation cases,  one cannot simply compare the snapshots at the same time stamp in different simulations. Instead, we must find the snapshots that represent the same {\emph evolution stage} in the photosphere. Given the time it takes the flux bundle to rise through the domain (see \tabref{tab:emer}), $t{=}24$\,h in the $98\times8$ implies an evolution time of about 18 hours after deducting the travel time of abount 6 hours. Therefore, the corresponding time stamp for the other simulation can be roughly estimated by the 
\begin{equation}\label{equ:t_stamp}
18\times \frac{f_{t}}{(f_{t})_{98\times8}}+\tau_{\rm travel},
\end{equation}
where $\tau_{\rm travel}$ can be found in \tabref{tab:emer} and values of $f_{t}$  are listed in \tabref{tab:list}. \figref{fig:btau1_multi} shows the vertical magnetic field in the $\tau{=1}$ surface at $t{=}24$\,h in the $98\times8$ run, $t{=}56$\,h in the $98\times18$ run, $t{=}56$\,h in the $196\times18$ run, and $t{=}81$\,h in the $196\times32$ run. We find that the asymmetry in the magnetic field, namely a coherent and strong leading spot and one or several fragmented flux concentrations in the following polarity, is consistent through all cases. Because the appearance of flux concentrations in the continuum intensity is primarily determined by their magnetic properties, the continuum intensity images for other simulation runs (not shown here) show a similar morphological asymmetry as in \figref{fig:asymm}(b).

We note that the leading spot in the $98\times18$ run has bifurcated into two flux concentrations at the time shown in \figref{fig:btau1_multi}(b), albeit, each of them is still a very coherent and reaches a peak field strength above 3\,kG. Despite that, the asymmetric properties found in the $98\times8$, $196\times18$, and $196\times32$ cases \footnote{The $196\times8$ case is not included in this analysis. However, it behaves very similar to the $98\times8$ case in all aspects.} are very pronounced  and consistent.

\subsubsection{Asymmetric properties in the magnetic structures beneath the photosphere}\label{sec:asym_vert}

\begin{figure*}[t!]
\center
\includegraphics{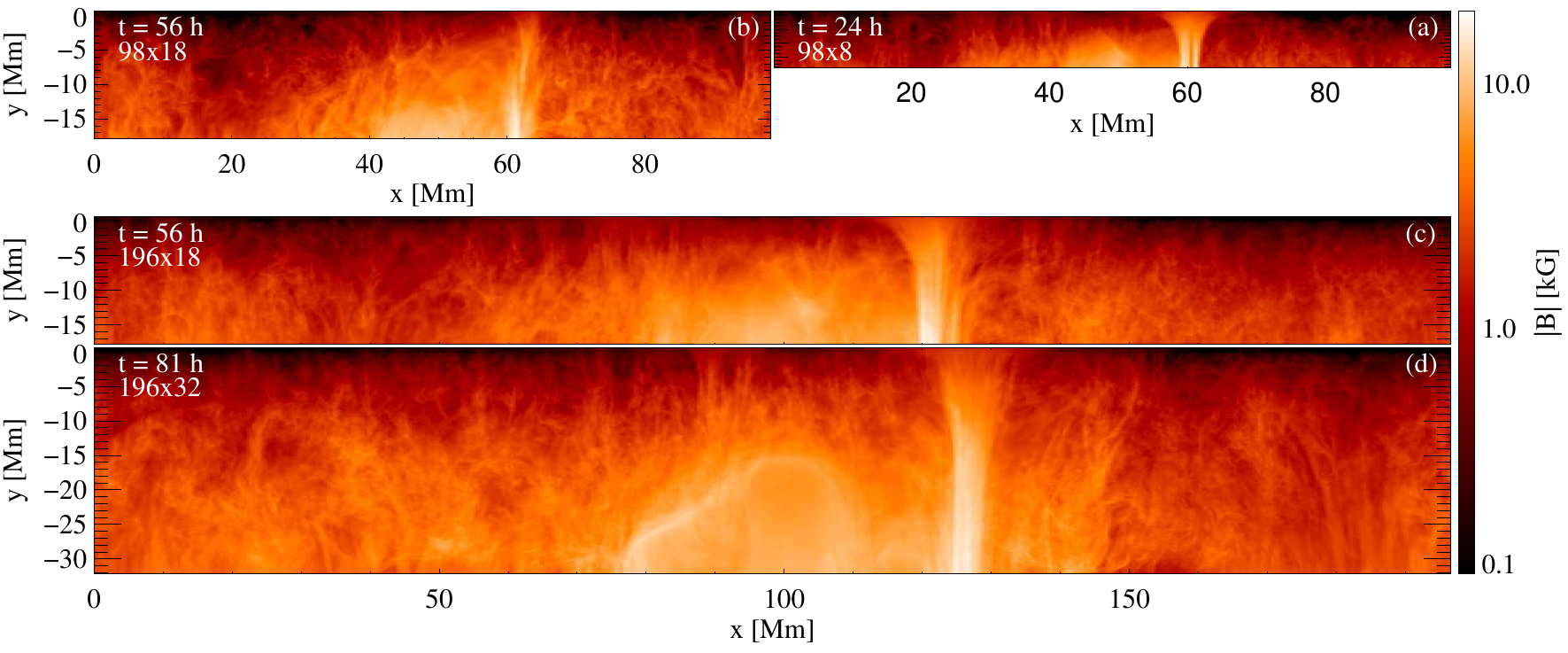}
\caption{Magnetic field strength ($\left | \mathbf{B} \right |$) in a vertical slices ($x-z$ plane) through the center of the domain ($y{=}49$\,Mm for 98\,Mm wide domain and $y{=}98$\,Mm for 196\,Mm wide domain) in four simulation cases. Times are identical to those in \figref{fig:btau1_multi}, which are selected to represent the same evolution stage. \label{fig:bvert_multi}}
\end{figure*}

The magnetic asymmetry of the sunspots is also present in the magnetic structure beneath the surface. The time stamps of \figref{fig:bvert}(b)(c)(d) showing the magnetic field strength in the central vertical slice are identical to the three times selected in \figref{fig:asymm} to illustrate the asymmetry in the photosphere. As shown in \figref{fig:bvert}, the leading polarity field has given rise to a coherent sunspot, when the following polarity field has not formed any coherent structure, which is consistent with the appearance in the photosphere. In addition, the emerging flux bundle display a very asymmetric shape, due the significantly stronger vertical magnetic field at the leading side. At $t{=}32$\,  the following spot also becomes visible in this snapshot, although it is much more fragmented and of clearly weaker magnetic field compared with the leading spot. To summarize, the magnetic structures in the convection zone actually present asymmetric properties similar to those observed in the photosphere.

We also examine the asymmetry of the magnetic field beneath the photosphere in other simulations as we have done for the asymmetry in the photosphere. For this purpose, we take the average of the magnetic field strength in the $y$ direction between the two dashed lines in each panel in \figref{fig:btau1_multi}. This is because a vertical slice through the center of the domain as in \figref{fig:bvert} cannot capture the sunspot pair and the flux bundle very well, if they are somehow away from the center $x-z$ plane, and the average in the $y$ direction gives a much better representation of the magnetic structure of the whole sunspot pair. 

\figref{fig:bvert_multi} shows the magnetic field strength beneath the surface at the same time instants as in \figref{fig:btau1_multi}. In each simulation case, we find a coherent and strong spot at the leading side of the emerging flux bundle, and there is basically no coherent magnetic structure at the following side. Consequently, as we have seen in \figref{fig:bvert}(c), the emerging flux bundle seems to incline forward (i.e., to the leading side), which reveals the much stronger vertical magnetic field at the leading side. The comparison confirms that both the sunspot asymmetry seen in the photosphere as well as that beneath the photosphere are consistent in the different simulation setups. And hence the asymmetric properties of the sunspots found in the simulations in this study is robust.

\subsubsection{Connection between the photosphere and the convection zone}
As presented in \sectref{sec:res_deep}, the leading spot in the $196\times32$ run is a monolithic magnetic structure anchoring in the downflow lane of the giant convective cell in the convection zone. The comparison in \figref{fig:bvert_multi} suggests that the same scenario is also valid for the other simulation cases. A clear implication of this scenario is that the positions of the photospheric spots are actually controlled by the downdraft lanes in which the sunspots anchor. This would further imply that the tilt angle of the sunspot pair in the photosphere essentially reflects that of the flux bundle giving rise to the sunspots. For instance, \figref{fig:asymm}(f)(i) provide a clear illustration of this relation in the $98\times8$ run. \citet{Fan+Fang:2014} showed that the tilt angle of emerging flux bundles generated in the convective dynamo are statistically consistent with observed mean tilt \citep[e.g.,][]{Stenflo+Kosovichev:2012}. The simulations presented in this paper make this comparison more straightforward, and confirm that the bipolar sunspot pairs formed by the emerging flux bundle in the convective dynamo model have title angles consistent with observations.

The downdraft in the convection zone also quite noticeably influences the shape of sunspots in the photosphere, which can be clearly seen in particular after the sunspots have formed, e.g., in \figref{fig:asymm}(f)(i). We find that the downdraft lane at the leading side of the giant convective cells has a compact patch of downflow at ($x{\approx}60$\,Mm, $y{\approx}50$\,Mm) that is clearly faster than surrounding downflows. In contrast, the downflow at the following side appears as an arch of similar flow speed lying between ($x{\approx}35$\,Mm, $y{\approx}40$\,Mm) and ($x{\approx}50$\,Mm, $y{\approx}60$\,Mm). Consequently the leading spot controlled by the compact fast downflow has coherent roundish shape, while the following spot follows the arch-like downflow pattern.

\subsection{Origin of the asymmetry}\label{sec:asym_ori}
After demonstration of the asymmetric properties of bipolar sunspot pairs, the question of how the asymmetries are generated remains. In the last section we have shown that the behavior of the photospheric sunspots, as well as their counterpart in the convection zone, are eventually determined by the magnetic and flow fields imposed at the bottom boundary. This implies that the asymmetric properties of the photospheric sunspots is a heritage of the asymmetries in the flow and magnetic field in the convection zone. 

\subsubsection{Asymmetries in the magnetic and flow field in the convection zone}
\begin{figure*}[t!]
\center
\includegraphics{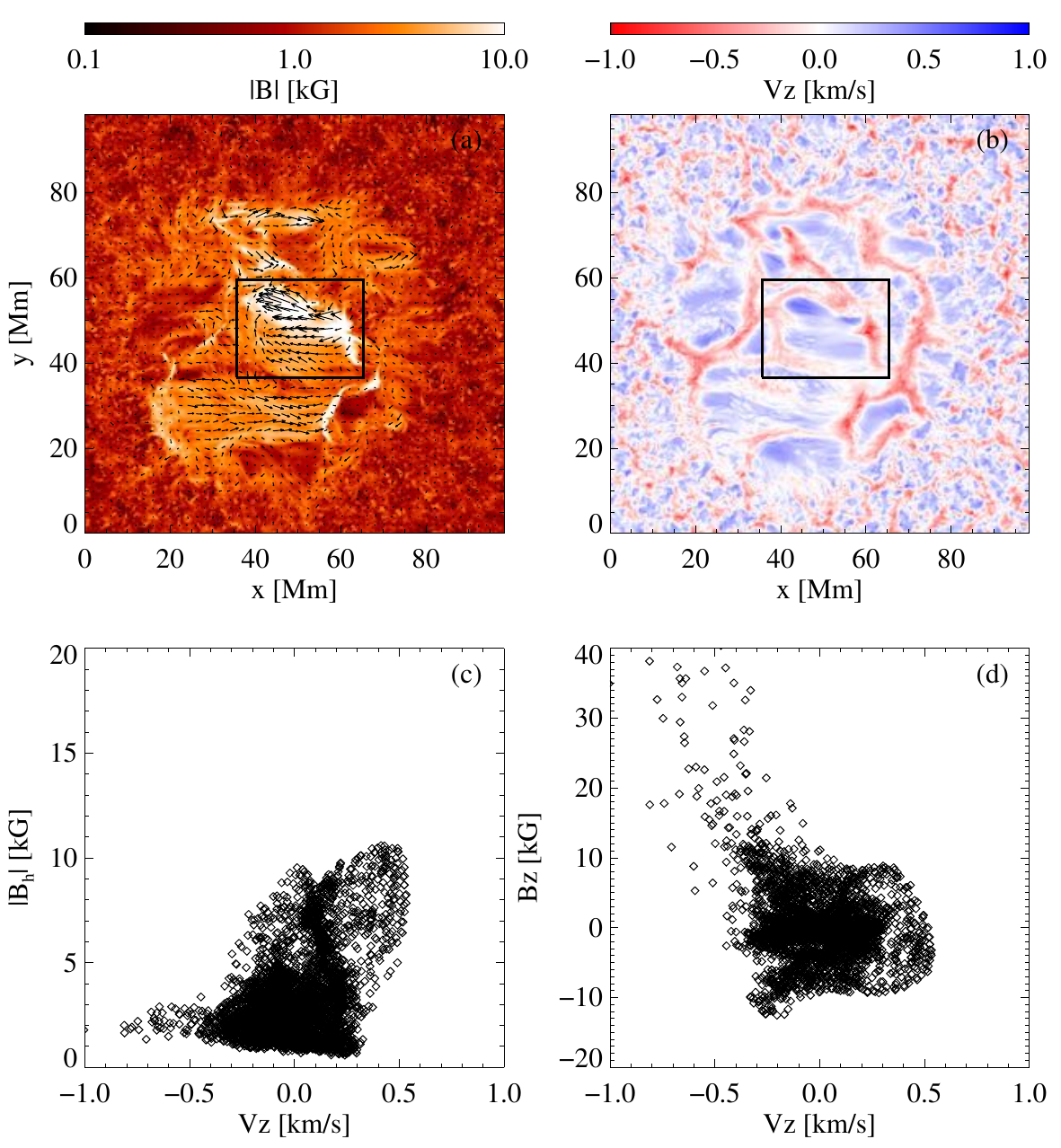}
\caption{Magnetic and flow field averaged between $t{=}24$\,h and $t{=}32$\,h at $z{=}-7$\,Mm. (a): Magnetic field strength, with arrows showing the horizontal magnetic field. (b) Vertical velocity. The black frame in (a) and (b) marks the area covering the central flux bundle that gives rise to Spot C1 and C2. (c): Relation between the strength of horizontal magnetic field and the vertical velocity in the marked region. (d): Relation between the vertical magnetic field and the vertical velocity in the marked region. \label{fig:vzbz}}
\end{figure*}

To better highlight the persistent asymmetries in the flow and magnetic fields near the bottom boundary, we calculate an average of the magnetic field strength and vertical velocity at $z{=}-7$\,Mm\footnote{This layer is very close the bottom boundary and represent the data imposed at the bottom boundary sufficiently well.} in a time period of 8 hours, and display the results in \figref{fig:vzbz}(a) and (b), respectively. The arrows on the field strength map shown in \figref{fig:vzbz}(a) indicate the horizontal component of the magnetic field.

The large-scale magnetic field and vertical flow averaged over 8 hours are very similar to that shown in \figref{fig:bvslices}, because they reflect the data imposed at the bottom boundary that evolve in a much longer time scale than the convection at this depth. We can see the central flux bundle embedded in a giant convection cell. The bulk of the flux bundle is mostly horizontal and cospatial with the upflow. The flux bundle is bounded by the downflow lanes around the convection cell. 

We use a black frame in \figref{fig:vzbz}(a) and (b) to highlight the area that covers the central flux bundle that gives rise to the most prominent sunspot pair. \figref{fig:vzbz}(c) displays a scatter plot of the strength of the horizontal magnetic field vs the vertical velocity in the regions marked by the frame. The lower part of the plot ($\left |\mathbf{B}_{\rm h}\right |< 5$\,kG) is dominated by magnetic field that is maintained by the small-scale dynamo. The most interesting feature is a branch that shows a field strength around 10\,kG and an upflow of  a few hundred \,m\,s$^{-1}$. It corresponds to the horizontal (toroidal in the original convective dynamo simulation) flux bundles emerging with convective upflows.

We also plot in \figref{fig:vzbz}(d) the relation of the vertical magnetic field to the vertical velocity in the region marked by the frame. The long branch to the upper-left of the panel corresponds to the foot of the flux bundle at the leading side (i.e., the leading spot), and the very short branch to the lower-left corresponds to the following spot. It clearly shows that the leading spot has a much stronger vertical field strength, and the same asymmetry is present in the downdraft lanes at which the sunspots anchor. 

\subsubsection{What creates the asymmetry in the magnetic field?}\label{sec:asym_mag}
In the following we give an explanation on the origin of the stronger vertical magnetic field at the leading side derived based the motion of fluid and induction of magnetic field. The convective motion in the Sun (except that in the most surface layers) behaves like an anelastic flow, because the sound speed is significantly larger than the typical flow speed. In this case, the continuity equation reads
\begin{equation}
\nabla \cdot (\rho \mathbf{v}) = 0, 
\end{equation}
which can be written as
\begin{equation}
\mathbf{v}_{h}\cdot\nabla_{h}\rho + v_{z}\frac{\partial \rho}{\partial z} + \rho \nabla \cdot \mathbf{v} = 0,
\end{equation}
where $\mathbf{v}_{h}$ and $v_{z}$ are the horizontal and vertical components of velocity, $\mathbf{v}$, respectively. $\nabla_{h}$ represents the operator of derivative on the horizontal components, i.e., $(\partial/\partial x, \partial/\partial y)$. Because the Sun is highly stratified, and hence the horizontal fluctuation of the density is negligible compared with the vertical variation, we can obtain
\begin{equation}
\frac{v_{z}}{H_{\rho}} = \nabla \cdot \mathbf{v},
\end{equation}
where $H_{\rho}{=-}\rho\mathrm{d}z/\mathrm{d}\rho$ is known as the density scale height. It reveals a simple but non-trivial relation between a converging flow (i.e., negative divergence) and a down flow (i.e., negative $v_{z}$).

The induction of magnetic field in ideal MHD reads
\begin{equation}
\frac{\partial \mathbf{B}}{\partial t} = \nabla \times (\mathbf{v} \times \mathbf{B}).
\end{equation}
We can reformulate the induction equation with vector identities, such that
\begin{equation}
\frac{\partial \mathbf{B}}{\partial t} + (\mathbf{v} \cdot \nabla)\mathbf{B} = -(\nabla \cdot \mathbf{v})\mathbf{B} + (\mathbf{B} \cdot \nabla)\mathbf{v} + (\nabla \cdot \mathbf{B})\mathbf{v},
\end{equation}
where the last term vanishes as $\mathbf{B}$ is solenoidal. Then the $z$ component reads
\begin{equation}
\frac{\partial B_{z}}{\partial t} + (\mathbf{v} \cdot \nabla)B_{z}  = -(\nabla \cdot \mathbf{v})B_{z} + (\mathbf{B} \cdot \nabla)v_{z}.
\end{equation}

The left hand side represents the temporal change of $B_{z}$ in a Lagrangian fluid element. The right hand side shows that vertical magnetic field can be intensified by a converging flow acting on the vertical magnetic field and/or a stretching/shear of the magnetic field by (the gradient of) a vertical flow. Considering a flux bundle in the $x$ direction, which is approximately the case in our simulations, we can further simplify the equation as
\begin{equation}
\frac{D B_{z}}{D t} \approx -\frac{v_{z}}{H_{\rho}}B_{z} + B_{x}\frac{\partial v_{z}}{\partial x}.
\end{equation}
The first term on the right hand side explains the relation of $B_{z}$ and $v_{z}$ as shown in \figref{fig:vzbz}(d). The stronger vertical field strength at the leading side is apparently related to a faster downflow, but physically resulted from the converging flow induced by the downflow. 

Furthermore, the contribution of the latter shear term is also expected to be significant, in particular when the emerging flux bundle was formed in the solar convective dynamo. As shown in \figref{fig:emg3d}, the emerging flux bundle is significantly sheared downward by the strong downflow lane on its leading side and forms a U-turn, however, the following side of the flux bundle is only slightly pressed downward by a much weaker downflow.  

To summarize, the stronger downflow on the leading side of the giant convection cell is responsible for the stronger and more vertical magnetic field at the same location. This magnetic structure eventually develops into the stronger and more coherent leading spot in a bipolar sunspot pair. Although the demonstration and analysis of the asymmetry properties of sunspots are more based the $98\times8$ run, we emphasize that the asymmetry properties and relevant physical processes can be consistently found in other simulation cases.

\subsubsection{Origin of the asymmetrical flow in giant convection cells}
After highlighting that the flow field imposed from the global dynamo simulation drives the formation of asymmetric magnetic field in the emerging flux bundle, the remaining question is what gives rise to the asymmetry of the flow field in the convection zone. This is beyond the scope of the simulation presented in this paper, but has been address in convective dynamo models. \citet{Fan+Fang:2014} showed that the giant convection cell, in which the emerging flux bundle embeds, moves prograde in respect to other features at the same latitude and depth. As a result, the leading polarity head of the flux bundle is pushed closer and harder into the downdraft lane (at its leading side) and the following polarity head is dragged away from the downdraft lane at the following side. In order to transport the excessive mass pushed into the leading side downdraft lane by the prograde motion, the downflow speed at the leading side of the convection cell naturally becomes higher. Then the mechanism we discussed in previous sections becomes in action, and gives rise to the asymmetry properties that are inherited by bipolar sunspot pairs in the photosphere. 

The prograde motion is a common and natural consequence in convection at low latitude in a rotating spherical shell. It can be considered as a systematic tilt of convection cells towards the direction of rotation, which is actually a representation of the positive correlation between the radial velocity and azimuthal velocity, as demonstrated by \citet{Busse:2002} and \citet{Aurnou+al:2007}. The same behavior is also found in the anelastic simulations by \citet{Featherstone+Miesch:2015}, in which a density stratification is taken into account to better represent the solar convection zone. In the authors words, they suggested that the tilt is produced by ``the combination action of the spherical geometry, the density stratification, and positive nonlinear feedback from the differential rotation they (i.e., {\it banana cells}, see \citealt{Miesch:2005}) establish". Moreover, they found that the correlation between the radial and azimuthal velocity leads to an outward angular momentum transport that plays a crucial role in the differential rotation and meridional circulation established in their simulations.

\section{Summary and Discussion}\label{sec:sum}
In this paper we present a series of simulations based on the method shown in \citet{Cheung+al:2010} and \citet{Rempel+Cheung:2014}. The major advantage of the simulations in this study is that we use the magnetic and flow fields adapted from a solar convective dynamo simulation \citep{Fan+Fang:2014} as the input for the fully compressible realistic MHD simulations of the upper convection zone. This setup further improves the degree of the realism of flux emergence simulations and leads to formation of the sunspots and active region more similar to those on the real Sun. The coupling between models for different layers of the convection zone brings us a unique opportunity to study the generation of emerging flux bundles in the convection zone and how they emerge into the surface. In particular,  the coupled model allows us to address the origin of the asymmetries in observed bipolar sunspot pairs. The main results are summarized as following.

1. The flux emerged to the surface of the Sun leads to formation of sunspots. The flux bundles emerge rather coherently before reaching the uppermost few Mm of the convection zone. The coherent flux bundles fragment into small magnetic elements in sizes similar to that of granules. The small magnetic elements rise to the photosphere and coalesce to form large flux concentration. 

2. Penumbral filaments form spontaneously as the magnetic flux concentration develops from a pore to a sunspot. Most interestingly, real-looking penumbra forms in our simulation without any approach to enforce the inclination of the magnetic field lines around the umbra. A large section of the penumbra is dominated by radial inflows, which is opposite to the direction of the typical Evershed effect in stable sunspot penumbrae.

3. The prominent sunspot in our simulations is a rather monolithic structure that anchors in the downdraft lane of a giant convection cell at up to 32\,Mm beneath the photosphere. We find a dominating downflow inside the sunspot deeper than 15\,Mm beneath the surface, while in the upper 15\,Mm the flow shows a mixture of up and downflows. The flow field outside the sunspot is dominated by a radial inflow below $z{\approx}-15$\,Mm and a radial outflow between $z{\approx}-15$\,Mm and the surface, both of which extend more than 30\,Mm in the radial direction.

4. Simulated sunspot pairs reproduce asymmetries in bipolar sunspot pairs observed on the Sun. The apparent cause of the asymmetries is the stronger vertical magnetic field, not only at the surface but also in the convection zone, of the leading spots. The giant convection cell, in which the emerging flux bundle is embed, move prograde in respect to its surrounding, and pushes the flux bundle into the downdraft lane at its leading side. This induces stronger downflow at the leading side of the convection cell and consequently gives rise to the stronger vertical magnetic field in the leading polarity. The prograde motion of flux bundles is the key of the origin of sunspot asymmetries and are essentially established when they are generated in the convection zone \citep{Fan+Fang:2014}. This behavior is a natural consequence of giant cell convection in a rotating shell.

\subsection{On the flux emergence process}
The behavior of emerging flux bundles found in our simulations is consistent with what was described by \citet{Cheung+al:2010} and was already noticed earlier in observations \citep[e.g.,][]{Zwann:1985}. \citet{Cheung+al:2010} suggested that the fragmentation of coherent magnetic structures is an efficient way to remove the mass load, and allows the magnetic field to rise further. The result was confirmed by \citet{Rempel+Cheung:2014} with differently configured flux tubes and a significantly deeper domain (i.e., larger density contrast). \citet{Stein+Nordlund:2012} studied the emergence of uniform, untwisted, horizontal magnetic field from 20\,Mm below the surface. The process of flux emergence in the photosphere is generally inline with what \citet{Cheung+al:2010} found.   

The simulations in the present paper employ a more realistic setup of input magnetic and flow fields, which is radically different from the analytically defined semi-torus flux tube or uniform field used in aforementioned works. We also vary the amount of magnetic flux by a factor of 4, and test an unprecedented density contrast (in the $196\times32$ run) in realistic simulations flux emergence, and conduct an experiment with significantly higher resolution. All simulation runs show results that are generally consistent with previous works. Therefore, it suggests that the process of flux emergence and formation of sunspots found in these works is robust.

In the simulations presented in this paper, the mean speed of upflows that advect the emerging flux bundles into the domain is similar to the mean convective upflow at the bottom boundary. In this setup, we find that the rising speed of flux bundles is determined by the mean convective upflow speed, which is similar to what \citet{Stein+al:2011} found in the weak field ($\vert \mathbf{B}\vert=10^4$\,G) case. The consistency of results is expected, because the field strength in the emerging flux bundle in our simulation is very close to 10$^4$\,G and both simulations are based on similar principles, i.e., realistic simulations of the upper solar convection zone. 

The large-scale flows in the near-surface layer are not affected by emerging flux bundles, which is consistent with the inversion of the horizontal flow field during flux emergence by helioseismology \citet{Birch+al:2016}. The mostly unchanged mean upflow velocity during flux emergence is also in line with the conclusion from the study of the travel time of emerging flux bundles. These results provide a clear evidence that the emergence of magnetic flux bundles in these simulations is mostly controlled by the convective upflows. Furthermore, the convective upflow is expect to still play a vital role in the flux emergence process observed at the photosphere and may help people to understand the relation of the flux emergence rate to the total magnetic flux emerged (see e.g., \citealt{Otsuji+al:2011,Norton+al:2017}).

\subsection{On the formation of sunspots}
Coherent flux bundles break into to the photosphere as small-scale magnetic dipoles that seem to embed in granular cells. These granular cells are strongly distorted when opposite polarity field moves apart. This behavior can be clearly seen in \figref{fig:highres}. Strong flux concentrations are formed by coalescence of small-scale magnetic features in the same polarity. This is consistent with the results presented by \citet{Cheung+al:2010}, \citet{Rempel+Cheung:2014}, and \citet{Stein+Nordlund:2012}, despite of the very different configurations on the magnetic flux injected from the bottom boundary in these models.

We notice an intriguing difference in the formation process beneath the surface. For instance, Spot C1, the leading spot in the central sunspot pair, is already well formed as a coherent vertical magnetic structure in the convection zone before its surface part develops into a strong sunspot. Namely, the sunspot forms from bottom to top. However, the leading sunspot in simulations by \citet{Rempel+Cheung:2014} first forms at the surface, and the coherent magnetic structure in the convection zone seems to develop from the surface to the deeper convection zone, i.e., the sunspot forms from top to bottom. 

The most likely cause of this difference is the flow field associated with the magnetic structures.  In simulations presented in this paper, a sunspot is anchored at a downflow lane that can facilitate formation of vertical magnetic structures, as discussed in \sectref{sec:asym_mag}. In contrast, \citet{Rempel+Cheung:2014} imposed a retrograde flow along the flux tube, which is mostly an upflow at the leading footpoint of the flux tube. Such a flow pattern can prevent a coherent vertical flux tube from forming. It is worthy to note that the formation process we observe at the surface is still similar in both cases. This implies that both bottom-to-top and top-to-bottom scenarios are possible, however, observations in the photosphere might not be sufficient to determine the actual process of sunspot formation.

The models discussed above consider the process of flux emergence and sunspot formation in one complete picture, and demonstrate the importance of the interaction between the emerging magnetic field and the plasma flows. Leaving the flux emergence out, \citet{Kitiashvili+al:2010} showed that uniformly distributed field can be intensified and form a flux concentration corresponding to a pore. A series of works by \citet{Warnecke+al:2013,Warnecke+al:2016}, \citet{Brandenburg+al:2013}, and \citet{Kaepylae+al:2016} investigated the intensification of pre-imposed weak field by Negative Effective Magnetic Pressure Instability \citep[NEMPI,][]{Brandenburg+al:2011} in more idealized models. It is very likely that the effects found in this models are also present in more complex realistic simulations. And all effects that can facilitate formation of strong flux concentrations, such as radiation, ionization, stratification, and turbulence, are entangled in realistic models shown in this paper. However, it is very difficult to quantify their individual contribution.

\subsection{On penumbral structures}
The penumbra of a sunspot is believed to be a manifestation of strong and highly inclined magnetic field surrounding the sunspot umbra. Penumbral structures of sunspots and their dynamics have been studied with radiative MHD simulation for either a section of the sunspot \citep{Heinemann+al:2007,Rempel+al:2009} or a fully circular sunspot \citep{Rempel+al:2009.sci,Rempel:2011}. These modes successfully reproduced morphological and dynamical features of sunspot penumbra (see the review by \citealt{Rempel+Schlichenmaier:2011}, discussions in \citealt{Rempel:2012}, and reference therein). 

However, in simulations that successfully can reproduce penumbrae with extensive filaments comparable to observations, the strong horizontal magnetic around the sunspots umbrae needs to be enforced either by a special configuration as in \citet{Rempel+al:2009.sci} or by the top boundary as in \citet{Rempel:2012}. It is very intriguing that penumbral structures in the high resolution simulation in this paper show a decent length \emph{without} any artificially enforced field line inclination. Long penumbral filaments are more likely to appear in regions with ongoing flux emergence, for example, on the side to the opposite polarity spot. While on the side away from the opposite polarity spot we often find less and shorter penumbra filaments. This seems to indicate a close relation between the flux emergence and formation of the penumbral structures. In observations, penumbrae formation is also found to coincide with flux emergence\citep[e.g.,][]{Schlichenmaier+al:2012,Shimizu+al:2012,Zuccarello+al:2014,Kitai+al:2014}, in particular the emergence of small magnetic dipoles around sunspots. We also note that \citet{Schlichenmaier+al:2010} reported that penumbra first forms on the opposite side of flux emerging region.

The Evershed flow has been successfully reproduced in a series of simulations, and the robustness of the numerical results is also extensively tested by \citet{Rempel:2012}. A preliminary analysis shows that penumbral filaments in our simulation are dominated by radial inflows, although part of the penumbra still shows an outflow that is consistent with the Evershed effect. This may reflect the difference between a stable sunspot (penumbra) and a developing sunspot (penumbra), where flux emergence is still active. Detailed observations of the flows in forming penumbrae seem to be very rare. \citet{Leka+Skumanich:1998} and \citet{Murabito+al:2016} noted that the Evershed flow appears very quick after the penumbra is formed or when the penumbra is forming. In contrast, \citet{Schlichenmaier+al:2012} and \citet{Romano+al:2014} found an inflow to the pore before the penumbra forms, and the flow direction is eventually reversed into the normal Evershed flow. Since what they found in the early stage seems similar to the behavior of our simulation, it will interesting to evolve the high resolution simulation longer to examine if the normal Evershed flow would prevail over the inflow. The analysis of the penumbral structures and dynamics in this paper is still very preliminary, and a detailed study will be given in a following paper.

\subsection{On subsurface structure of sunspots}
The subsurface structure of sunspot and their effective anchoring depth are amongst most debated fundamental properties of sunspots \citep[see the discussion in Sect 3.1][and references therein]{Moradi+al:2010}. It was proposed decades ago that magnetic flux tubes giving rise to sunspots are generated at the base of the convection zone, and the photospheric sunspots are suggested to be eventually connected to the bottom of the convection zone. Only in recent years, a few solar convective dynamo simulations demonstrated that coherent emerging flux bundles can also be generated in the bulk of the turbulent convection zone. By coupling the dynamo simulation of \citet{Fan+Fang:2014} and a simulation of the upper-most layer of the convection zone, we are able to depict a full picture of sunspots.

We focused on the strongest sunspot in the $196\times32$ run for the study of subsurface structure and flow field. This spot appears as a rather monolithic and deep-anchored structure rooted in the convective downdraft lanes at 32\,Mm below the photosphere. This is the depth of the bottom boundary, and more importantly, the depth where we extracted the data from the convective dynamo simulation. To complete the picture, the magnetic field structure in the convection zone below a depth of 32\,Mm is further described in the convective dynamo simulation of \citet{Fan+Fang:2014}. They showed that a sunspot pair is formed by the emergence of the upper part of a hair-pin shaped flux bundle, while the lower part of the hair-pin is well mixed with other magnetic structure in the convection zone. Therefore, on one hand, these sunspots are clearly not a surface phenomenon. They neither disperse or fragment after a short distance below the surface, nor become dynamically disconnected with the magnetic field in the convection zone. On the other hand, they are not anchored at the base of the convection zone as assumed in rising thin flux tube models. 

The flow field in the vicinity of a sunspot is also an integral part of the sunspot structure. Studies have been carried out not only for stable sunspots \citep{Rempel:2011.sub,Rempel:2015}, but also for sunspots forming through flux emergence \citep{Rempel+Cheung:2014}. The simulations presented in this paper and aforementioned are based on generally the same MHD equations and numerical method. The most substantial difference between these simulations is how the magnetic field is introduced and the treatment of the bottom boundary condition.

The effective domain size for the azimuthal average shown in \sectref{sec:res_deep} is approximately 80\,Mm wide and 32\,Mm deep, which is considerably wider and deeper than those in most of previous simulations. In the near-surface layer of the convection zone outside the sunspot, we find that an upflow follows the expanding boundary of the sunspot and transients gradually to an large-scale outflow that extend for more than 30\,Mm. Similar flow patterns were also found in previous simulations, albeit the spatial scale of the flow seems to be related to the domain size.

 The result in our simulation confirms the robustness of the outflows found in previous simulations. The existence of these outflows is quite independent from the domain sizes of the simulations, which vary from 48\,Mm to 98\,Mm in width and from 8\,Mm to 32\,Mm in depth. Moreover, it is independent from the formation processes of the sunspot, because the sunspots in \citet{Rempel:2011.sub,Rempel:2015} are formed from pre-existing strong magnetic field set by the initial condition, while those in \citet{Rempel+Cheung:2014} and this paper are formed through emergence of the magnetic flux introduced through the bottom boundary. And for the same reason, the bottom boundary condition does not seem to influence the behavior of the outflow. 
 
In summary, it gives a strong indication that the driving mechanism of the outflow is a surface effect. This is consistent with the analysis in \citet{Rempel:2015} and this paper, which show that the primary cause of the outflow is the reduction of the downflow fill factor and the resulting vertical mass flux imbalance beneath the penumbral area of the sunspot. Furthermore, the outflow is present around a sunspot with a normal Evershed flow, a sunspot with a reversed Evershed flow, and a naked sunspot with no Evershed flow. It implies that this outflow in the vicinity of the a sunspot has an independent origin from the Evershed flow in the penumbra. 

\subsection{On the origin of sunspot asymmetry}
We find that the asymmetric properties of bipolar sunspot pairs in the photosphere actually originate from the intrinsic asymmetric properties of the emerging flux bundle generated in the solar convective dynamo.  This is in line with the expectation in pioneering studies of rising thin flux tubes \citep{Fan+al:1993,Caligari+al:1995} that the asymmetry in magnetic field strength eventually leads to the morphological asymmetry in photospheric sunspots. 

However, the asymmetry in the flow field and the origin of the magnetic field asymmetry are \emph{completely different} in modern models. When a flux tube generated at the base of the convection zone moves to the surface, a retrograde flow develops as a consequence of angular momentum conservation. Inspired by this, \citet{Rempel+Cheung:2014} simulated the emergence of flux tube with a retrograde flow. They found that, although the leading spot eventually becomes more coherent than the following spot, the leading spot forms {\emph after} the following spot, due to a strong diverging flow in the early stage that prevents the leading spot from forming. This behavior contradicts with the majority of bipolar sunspot pairs observed by \citet{McIntosh:1981}. 

In conclusion, we couple a solar convective dynamo simulation to realistic simulations of the upper most layers of the convection zone. The coupled model depicts a comprehensive picture of the generation of magnetic flux bundles in the convect zone, emergence of magnetic flux to the surface, and formation of sunspots. Furthermore, the bipolar sunspot pairs in the model self-consistently reproduce the systematic asymmetries of sunspots observed on the Sun. This model offers a plausible explanation on the origin of solar active regions and sunspots and sheds new light on the origin of their long-known asymmetries.

\acknowledgments
We thank the anonymous referee for helpful suggestions. The National Center for Atmospheric Research (NCAR) is sponsored by the National Science Foundation. We thank Dr. Mark Miesch for comments that improve this manuscript. F.C. is supported by the Advanced Study Program postdoctoral fellowship at NCAR. This work is also supported in part by the NASA LWSCSW grant NNX13AJ04A and NASA grant NNX14AI14G. Computing resources were provided by the NCAR's Computational and Information Systems Laboratory, sponsored by the National Science Foundation, on Yellowstone (http://n2t.net/ark:/85065/d7wd3xhc).

\bibliography{reference}

\end{document}